\documentclass[a4paper,11pt]{article}
\pdfoutput=1 

\usepackage{jinstpub} 

\usepackage{lineno}
\title{\boldmath First survey of centimeter-scale AC-LGAD strip sensors with a 120 GeV proton beam}

\usepackage{siunitx} 
\DeclareSIUnit\sq{\ensuremath{\Box}}                           

\usepackage[rawfloats=true]{floatrow}

\usepackage{subcaption}
\makeatletter
\renewcommand\p@subfigure{\thefigure.}                        
\makeatother

\usepackage{multirow}

\usepackage{xcolor}

\author[a,1]{Christopher Madrid%
\note{Corresponding author.},}\emailAdd{cmadrid@fnal.gov}
\author[a]{Ryan Heller,}
\author[d,e]{Claudio San Mart\'in, }
\author[g]{Shirsendu Nanda,}
\author[a]{Artur Apresyan,}
\author[d,e,f]{William K. Brooks,}
\author[b]{Wei Chen,}
\author[b]{Gabriele Giacomini,}
\author[h]{Ohannes Kamer Köseyan,}
\author[a]{Sergey Los,}
\author[a]{Cristi\'an Pe\~na,}
\author[d,e,i]{Ren\'e Rios,}
\author[b]{Alessandro Tricoli,}
\author[a,c]{Si Xie,}
\author[g]{Zhenyu Ye}

\affiliation[a]{Fermi National Accelerator Laboratory, \\PO Box 500, Batavia IL 60510-5011, USA}
\affiliation[b]{Brookhaven National Laboratory,\\Upton, 11973, NY, USA}
\affiliation[c]{California Institute of Technology,\\Pasadena, CA, USA}
\affiliation[d]{Departamento de F\'isica y Astronom\'ia, Universidad Técnica Federico Santa María, \\Valparaiso, Chile}
\affiliation[e]{Centro Cient\'ifico Tecnol\'ogico de Valpara\'iso-CCTVal, \\
Universidad T\'ecnica Federico Santa Mar\'ia, Casilla 110-V, Valpara\'iso, Chile}
\affiliation[f]{Millennium Institute for Subatomic Physics at the High-Energy Frontier (SAPHIR) of ANID,\\Fern\'andez Concha 700, Santiago, Chile}
\affiliation[g]{University of Illinois at Chicago,\\Chicago, IL 60607, USA}
\affiliation[h]{Department of Physics and Astronomy,\\The University of Iowa, Iowa City, Iowa, USA}
\affiliation[i]{Departamento de F\'isica, Universidad de La Serena, \\Benavente 980, La Serena, Chile}

\abstract{
We present the first beam test results with centimeter-scale AC-LGAD strip sensors, using the Fermilab Test Beam Facility and sensors manufactured by the Brookhaven National Laboratory. Sensors of this type are envisioned for applications that require large-area precision 4D tracking coverage with economical channel counts, including timing layers for the Electron Ion Collider (EIC), and space-based particle experiments. A survey of sensor designs is presented, with the aim of optimizing the electrode geometry for spatial resolution and timing performance. Several design considerations are discussed towards maintaining desirable signal characteristics with increasingly larger electrodes. The resolutions obtained with several prototypes are presented, reaching simultaneous \SI{18}{\um} and \SI{32}{\ps} resolutions from strips of \SI{1}{\centi\m} length and \SI{500}{\um} pitch. With only slight modifications, these sensors would be ideal candidates for a 4D timing layer at the EIC.

}


\begin{document}
\maketitle
\flushbottom

\section{Introduction}


The use of 4-dimensional (4D) trackers~\cite{4DTracking_WhitePaper} that provide precise spatial and timing information is critical for many proposed experiments ranging from high energy collider physics to astroparticle experiments in space~\cite{instruments5020020}, and their development represents a key technological challenge.
Tracking detectors capable of achieving 5--50~\si{\ps} timing resolution and 5--30~\si{\micro\m} position resolution are needed for many proposed future colliders including the FCC-hh~\cite{Sickling, Wulz277931111}, Muon colliders~\cite{MuonColliderForum2022, Bartosik_2020},  and the Electron--Ion Collider (EIC)~\cite{AbdulKhalek:2021gbh}.
A major breakthrough in 4D tracking detector technology in recent years has been the development of silicon based sensors such as AC-coupled Low Gain Avalanche Detectors (AC-LGAD)~\cite{ACLGADprocess, 8846722, RSD_NIM, firstAC}.
These sensors have been demonstrated to be capable of achieving 30~\si{\ps} time resolution and 5-30~\si{\micro\m} position resolution~\cite{Apresyan:2020ipp,Heller_2022,TORNAGO2021165319,OTT2023167541}.

In this paper, we explore, for the first time, AC-LGADs of centimeter-scale strip length and relatively coarse pitch, enabling coverage of large areas with fewer channels. Longer strip sensors with sparse readout may offer better cost and performance for applications where channel count or electrical power density should be economized. 
Thanks to the excellent signal to noise ratio in AC-LGADs, sparse readout can be exploited without significant degradation of spatial or time resolution.

We also continue to explore the impact of the electrode metal size on the overall performance.
Studies of these key detector design parameters are of particular importance for the development of detectors at EIC, which can benefit from particle identification capability made possible by a 4D time-of-flight layer with \SI{30}{\pico\second} and tens of microns resolution.

We reported first studies of the performance of AC-LGAD sensors exposed to particle beams in previous publications~\cite{Apresyan:2020ipp,Heller_2022,TORNAGO2021165319,OTT2023167541}, focusing on fine pitch sensors targeting ultra-fine spatial resolution on the order of \SI{5}{\micro\m}. 
In this paper, we expand those studies to include several new detector prototypes with larger area and coarser pitch produced by Brookhaven National Laboratory (BNL) and tested at the Fermilab (FNAL) Test Beam Facility (FTBF)~\cite{FTBF}, focusing on in-depth studies of their performance. 
Section~\ref{sec:sensor} presents a description of the AC-LGAD sensors, and the experimental setup at the FTBF is described in Section~\ref{sec:setup}. 
Experimental results are presented in Sections~\ref{sec:properties}--\ref{sec:timing}. 
These results include measurements of sensor signal properties, detection efficiency, and position and time resolution. 
Conclusions and outlook are presented in Section~\ref{sec:discussion}.

\section{The AC-LGAD sensors}\label{sec:sensor}
\begin{figure}[htp]
\centering
\includegraphics[width=0.6\textwidth]{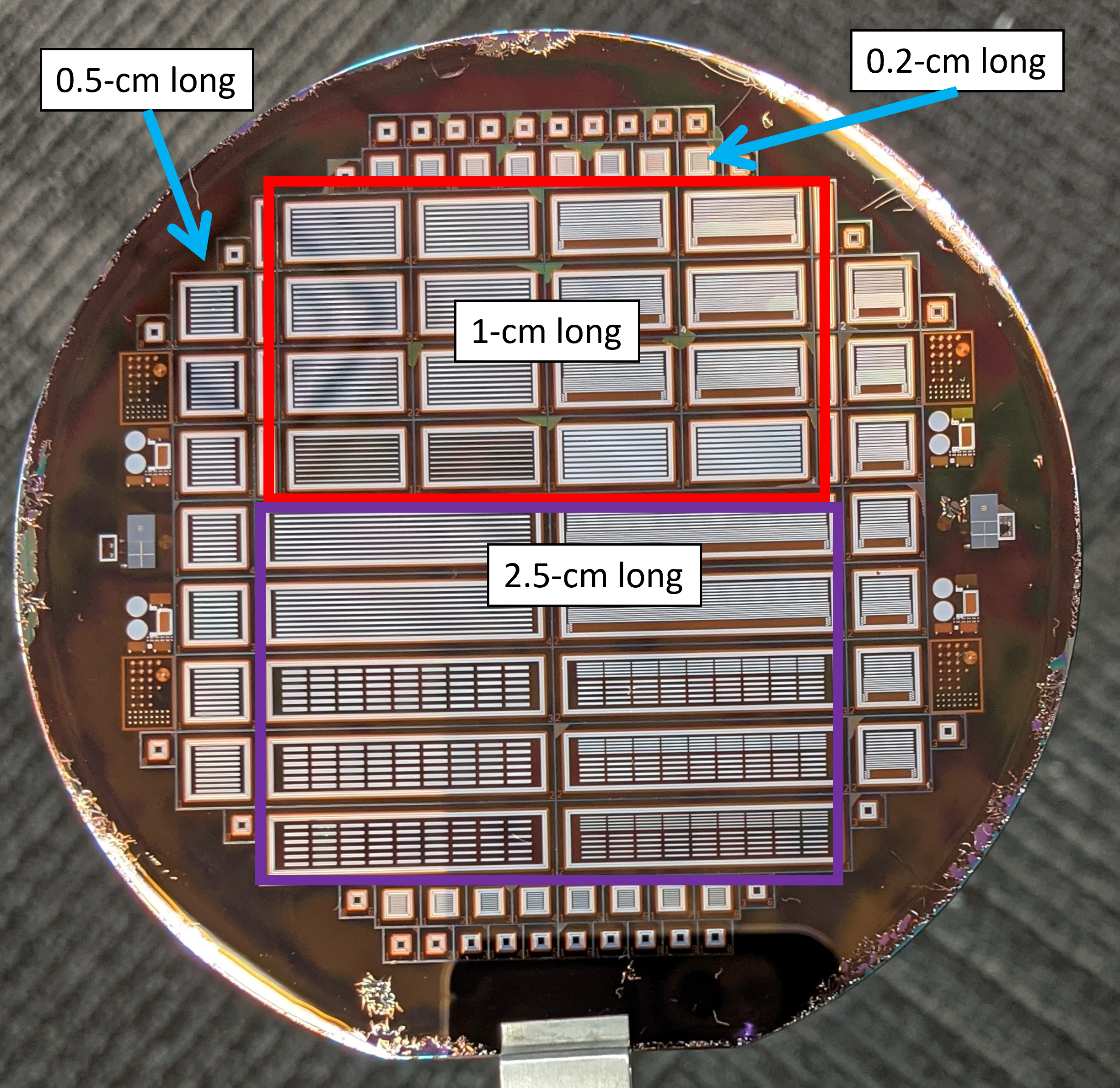}
\caption{Photograph of a 4~inch wafer fabricated with AC-LGAD strip sensors of different lengths. The wafer has 24 devices with \SI{1}{\cm} long strips, 10 devices with \SI{2.5}{\cm} long strips, and 14 devices with \SI{0.5}{\cm} long strips. 
\label{fig:wafer}}
\end{figure}

In this campaign, we exposed approximately 15 different AC-LGAD sensors to the FNAL proton beam. 
These were all produced in the same batch fabricated in a class-100 clean room at BNL on 4~inch, p-type epitaxial wafers. 
The epitaxial layer consists of high-resistivity silicon of thickness \SI{50}{\um}.
The substrate is a low-resistivity, \SI{300}{\um} thick wafer, serving as mechanical support, and its back is unpatterned.
A photograph of a wafer fabricated in this batch is shown in Figure~\ref{fig:wafer}. 
The wafer was populated with AC-LGAD strip sensors of active area $5\times 5$, $5\times 10$, and $5\times 25$~\si{\mm\squared}. 
The aluminum (about \SI{0.7}{\um} thick) for the AC-coupled strips is patterned on top of $\sim$~\SI{150}{\nm} of PECVD silicon oxide, which has been deposited over a uniform low-dose and high resistivity implant of phosphorus. 
At the border of the resistive n-layer, a junction-termination edge (JTE) is created by a deep phosphorus implantation. 
Embedded into the JTE, which runs all around the device, a high dose phosphorus implant is contacted by a metal frame, grounded during biasing of the sensor. 
The gain layer, a deep low-dose boron implant, is ion-implanted below the n-resistive layer and is \SI{20}{\um} from the JTE. 
The termination region, including guard rings, is the same for all devices.
The wafers have been passivated with polyamide while the full length of the metalized strip is exposed and available for wire-bonding.

For each strip length, a few variations in the pitch and width of the metal strips were produced:
\begin{itemize}
    \item \SI{5}{\mm} long strips: pitch/metal width 500/200 \si{\um} or a multi-pitch structure of pitch/metal width 300/150~\si{\um}, 200/100~\si{\um} and 100/50~\si{\um} into the same device;
    \item \SI{10}{\mm} long strips: pitch/metal width 500/100~\si{\um} or 500/200~\si{\um} or 500/300~\si{\um}, and the same multi-pitch series as in the \SI{5}{\mm} long strips;
    \item \SI{25}{\mm} long strips: pitch/metal width 500/200 and the same multi-pitch series as in the \SI{5}{\mm} long strips.
\end{itemize}

We focused the detailed analysis on a set of five sensors with large collected datasets that exhibit the characteristic behavior of this production. These sensors were selected to span the full range of strip lengths and widths produced, with the goal of extracting lessons for the geometric optimization of the electrodes. They all feature \SI{500}{\micro\m} pitch, as envisioned for an EIC timing layer. The parameters for these five sensors are shown in Table~\ref{table:SensorInfo} along with the operating voltages used. In the rest of this paper, we refer to them based on names that indicate the strip length, in \si{\milli\m}, and strip width, in \si{\micro\m}; for example, BNL 10-200 indicates the sensor with \SI{10}{\milli\m} length and \SI{200}{\micro\m} width.

\begin{figure}[htp]
\centering
\includegraphics[width=0.3\textwidth]{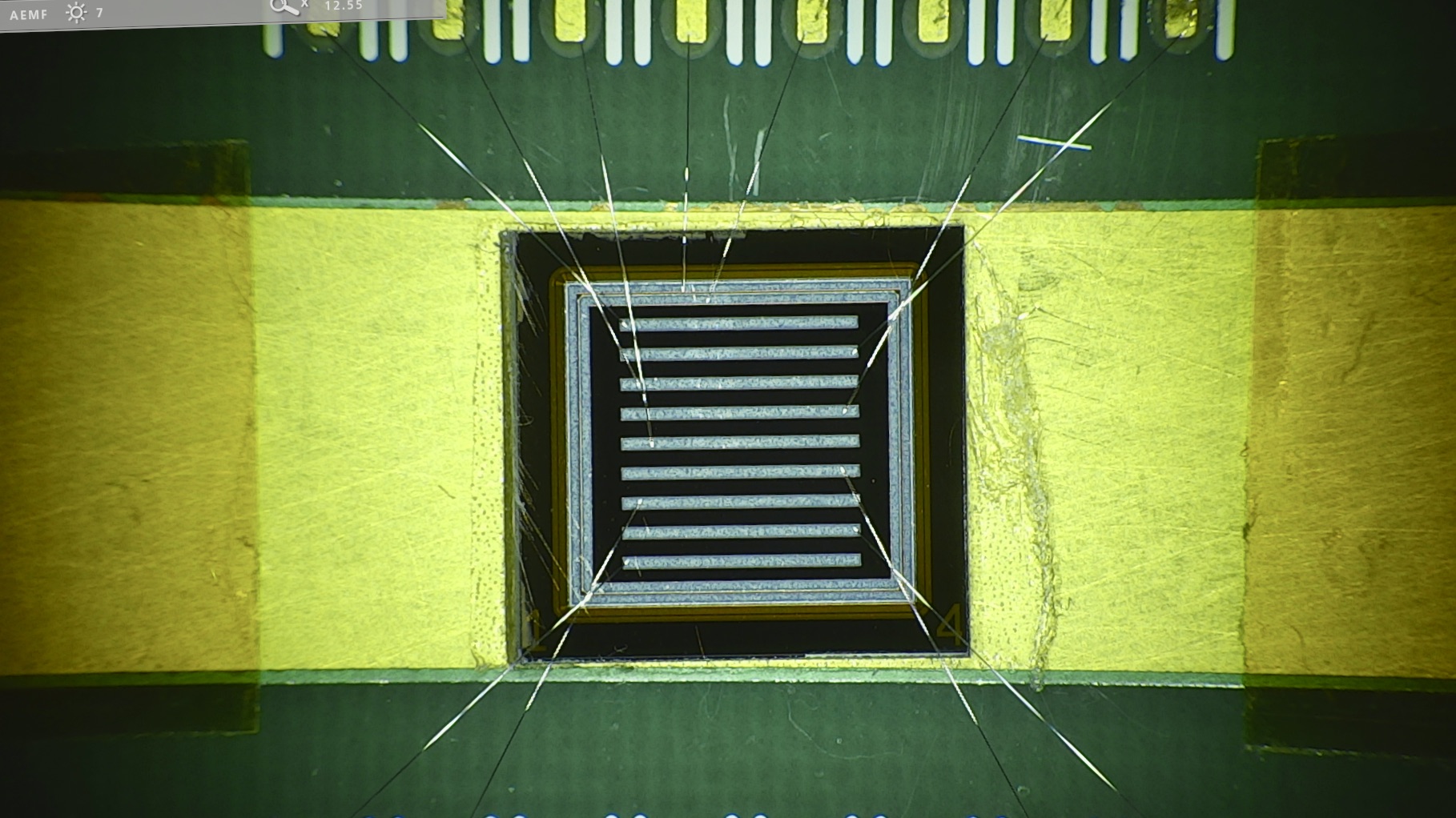}
\includegraphics[width=0.322\textwidth]{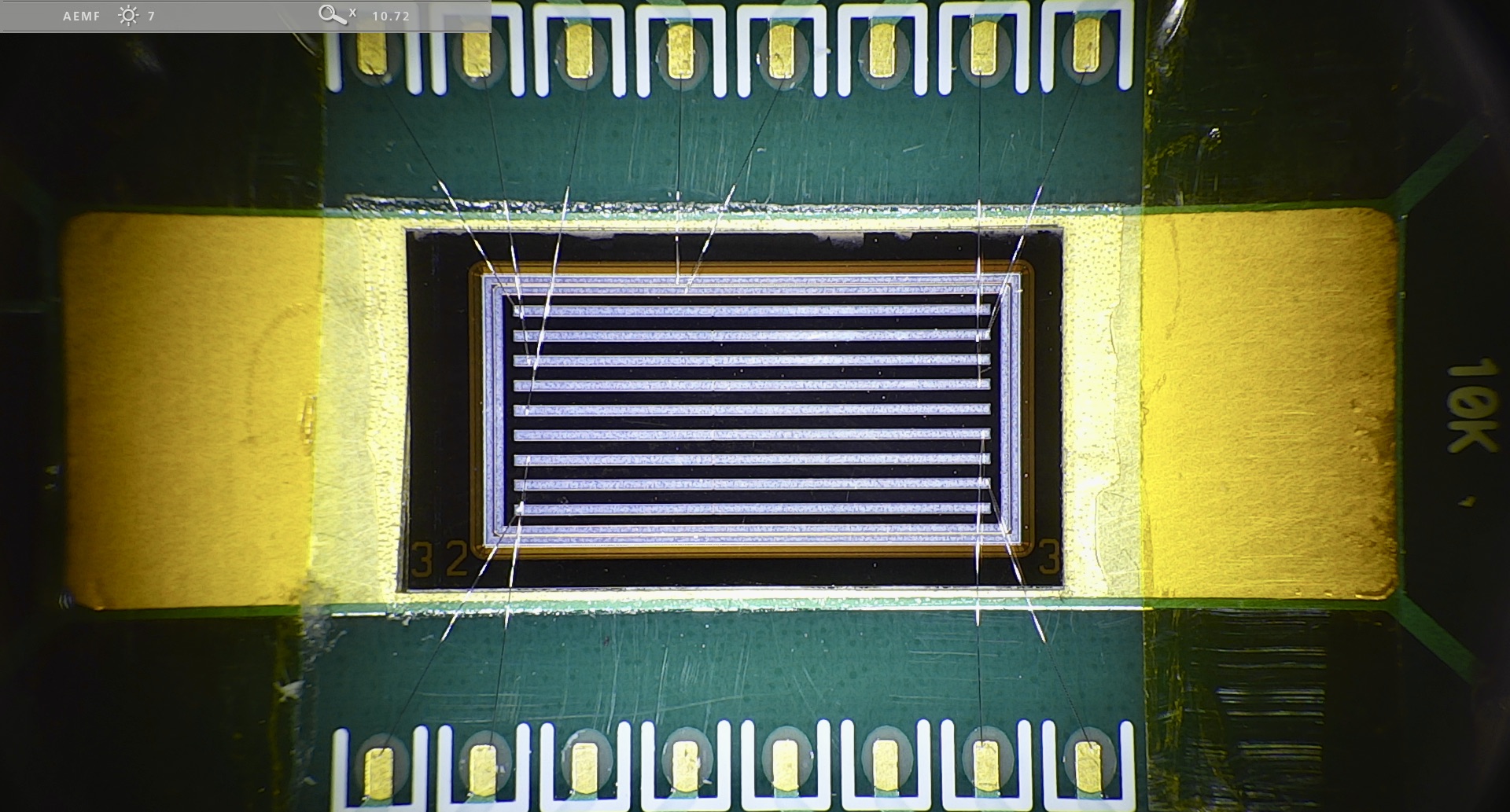}
\includegraphics[width=0.34\textwidth]{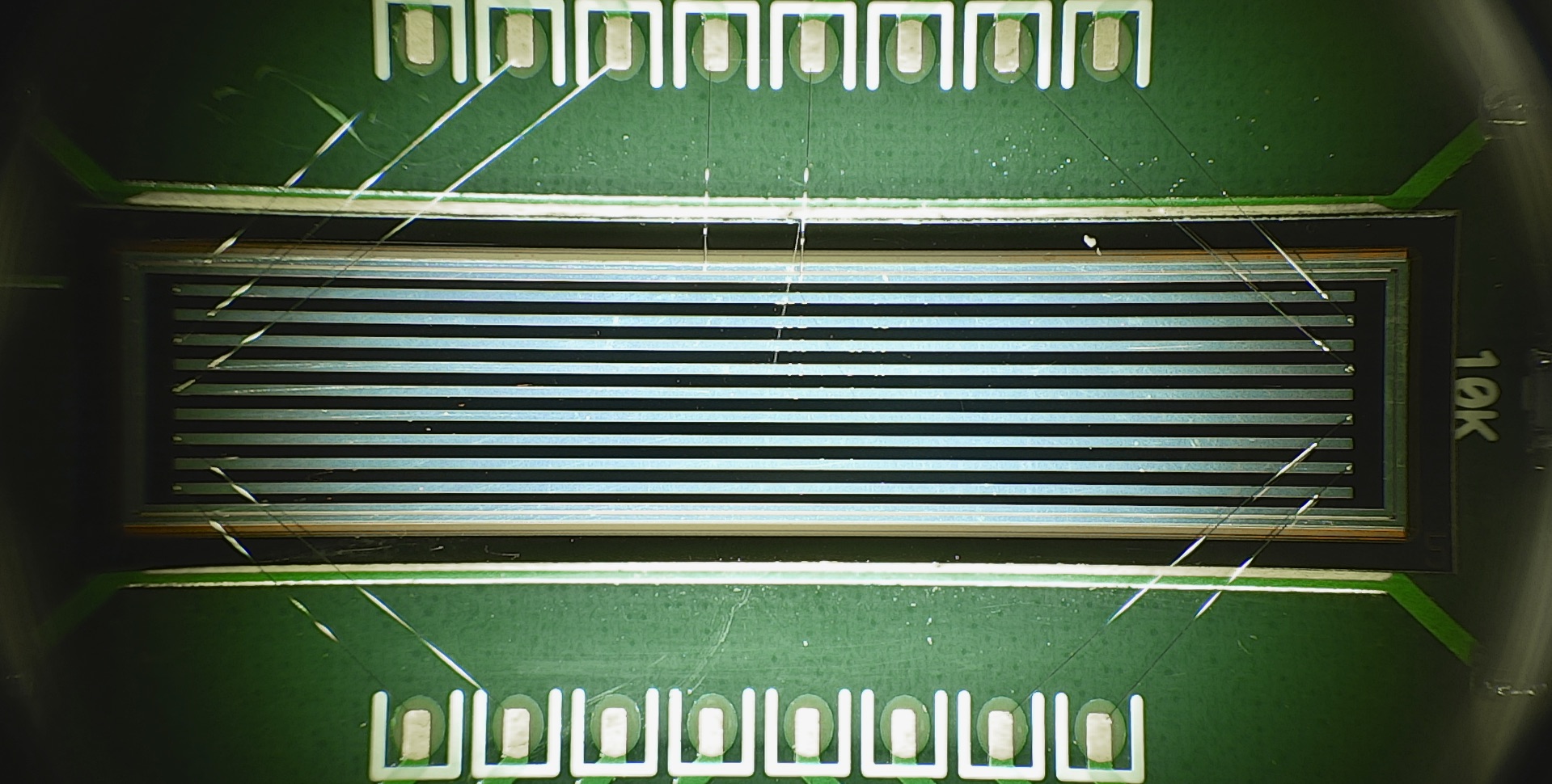}
\caption{The three strip length variations of BNL manufactured sensors tested at FNAL. BNL 5-200 (left), BNL 10-200 (center) and BNL 25-200 (right). The strips are read from alternating ends, to facilitate measuring of the longitudinal proton hit position (impact parameter) and compensating for signal propagation delays.
\label{fig:SensorImg}}
\end{figure}

Photographs of the three sensor length variations and wirebonding scheme are shown in Figures~\ref{fig:wafer} and ~\ref{fig:SensorImg}.

\begin{table}[htp]
  \centering
  \caption{A summary of the geometry and optimal operating voltage for the five sensors selected from the 15 sensor survey.}
  \begin{tabular}{ l | c | c | c | c | c}
  Name            & Pitch    & Metal Width & Length   & Thickness & Operating Voltage \\
  Unit            & \si{\um} & \si{\um}    & \si{\mm} & \si{\um}  & V      \\
  \hline\hline
  BNL  5--200     & 500      & 200         &  5       & 50        & 245    \\ \hline
  BNL 10--100     & 500      & 100         & 10       & 50        & 220    \\
  BNL 10--200     & 500      & 200         & 10       & 50        & 255    \\
  BNL 10--300     & 500      & 300         & 10       & 50        & 240    \\ \hline
  BNL 25--200     & 500      & 200         & 25       & 50        & 215    
  \end{tabular}
  \label{table:SensorInfo}
\end{table} 

\section{The experimental setup at the FNAL Test Beam Facility}\label{sec:setup}

\begin{figure}[htp]
\centering
\includegraphics[width=0.99\textwidth]{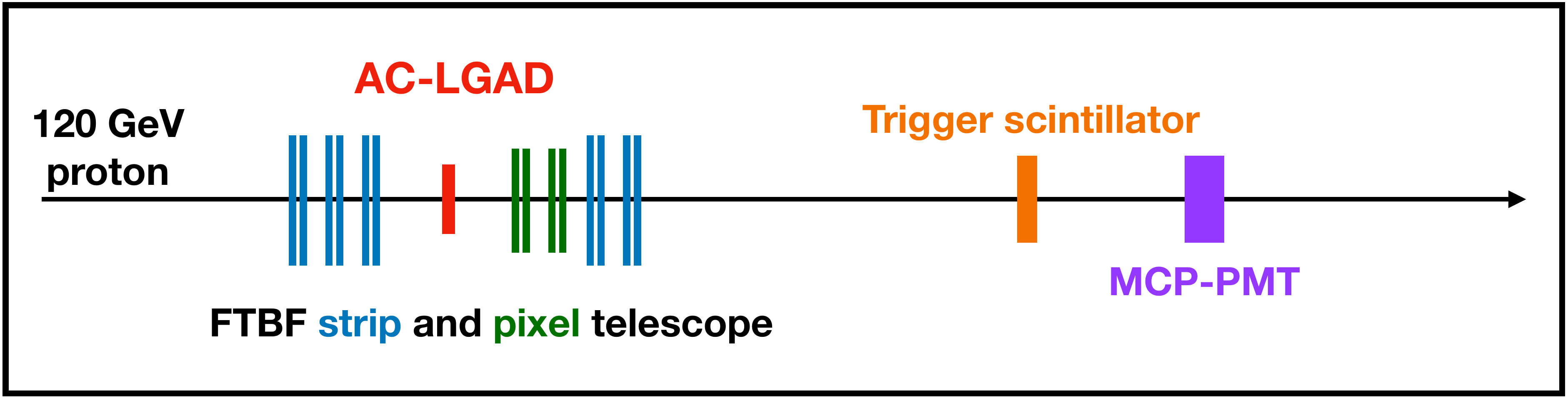}

\vspace{0.5cm}

\includegraphics[width=0.99\textwidth]{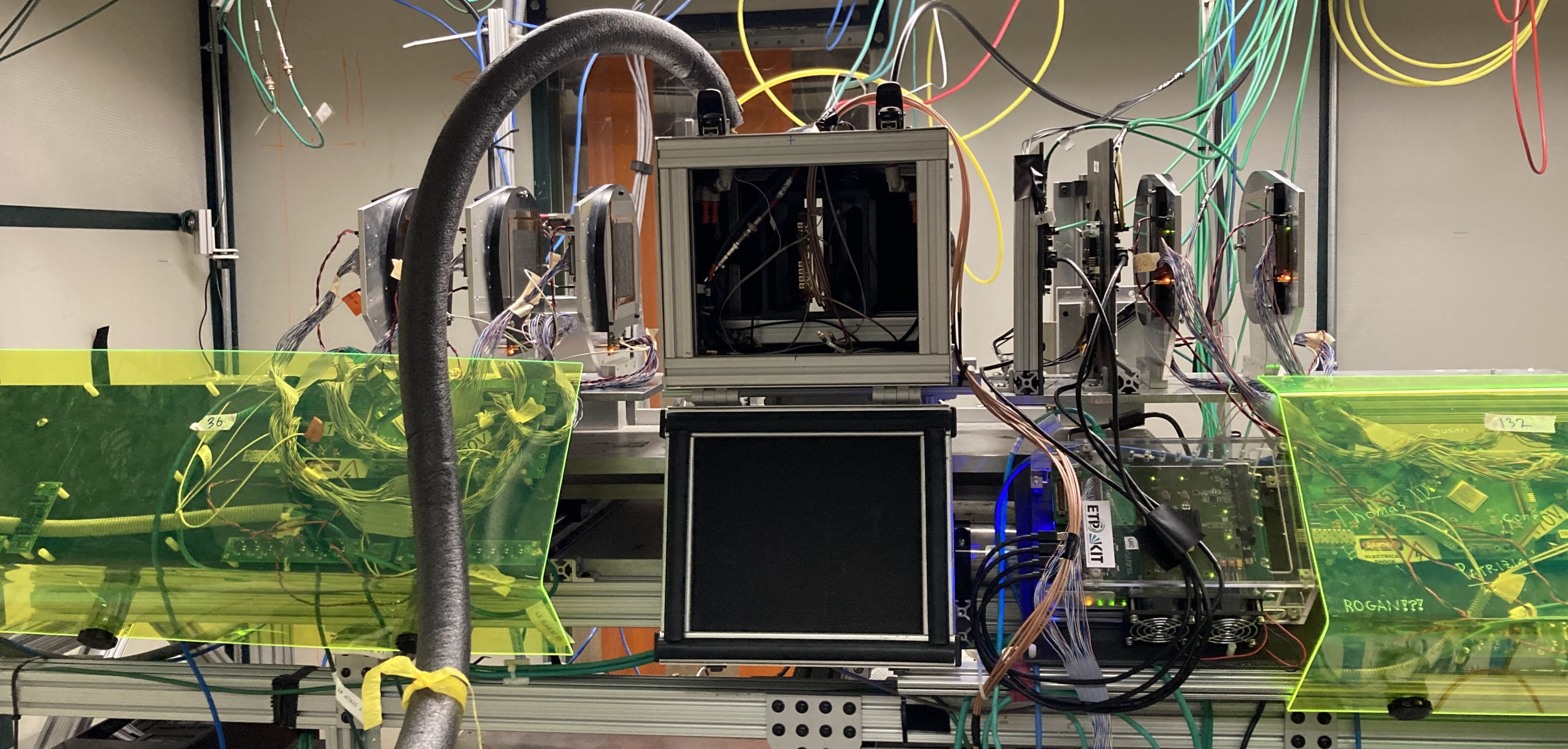}
\caption{Diagram of the AC-LGAD and reference instruments along the beamline (top). The environmental chamber placed within the FTBF silicon telescope (bottom). The telescope comprises five pairs of orthogonal strip layers and two pairs of pixel layers, for a total of up to 14 hits per track. 
\label{fig:FTBF_Box}}
\end{figure}

The results presented in this paper were collected at the Fermilab \SI{120}{\GeV} proton test beam facility, using the LGAD characterization setup described in detail in previous results~\cite{Apresyan:2020ipp,Heller_2022}. 
This setup is based on a silicon tracking telescope that measures the impact position of each proton, and a fast microchannel plate detector (MCP-PMT), which serves as a time reference with \SI{10}{\ps} resolution.
The AC-LGAD and MCP-PMT waveforms were recorded using an eight channel Lecroy Waverunner 8208HD oscilloscope with a bandwidth of \SI{2}{\giga \hertz} and a sampling rate of \SI{10}{GS/s} per channel. 
The AC-LGAD sensors were mounted on readout boards developed by Fermilab~\cite{HELLER2021165828}.
The readout boards are optimized for LGAD sensors and are capable of accommodating 16 channels. 
The trigger signal is generated by a scintillator detector located downstream from the AC-LGADs and distributed to the tracker and the oscilloscope. 
Figure~\ref{fig:FTBF_Box} shows a diagram of the setup along with an image of the environmental chamber that houses the AC-LGADs inside the telescope.

The pixel detectors inside the FTBF telescope have recently been upgraded from Phase 0 CMS pixel detectors with a pitch of $100\times150~$\si{\micro\m^2} to Phase 2 CMS pixel detectors with a pitch of $25\times100~$\si{\micro\m^2}. 
The new detectors use the RD53A chip~\cite{Garcia-Sciveres:2287593}. 
The new pixel detectors feature fast readout, which enables reliable track-finding even at high beam intensity. 
The new pixel detector modules also use a significantly improved data acquisition (DAQ), yielding increased operational robustness and enabling data-quality monitoring in real time. 
The active area of each pixel layer is $13.6\times9.6~$\si{\milli\m^2}, and they are aligned for an overlapping region of about $9.6\times9.6~$\si{\milli\m^2} with four expected pixel hits per track. 

Additionally, the procedure to align the sensors under test with the telescope was also improved. 
Previous measurements relied on sensor alignment that only considered variations in the sensor position along the beamline and rotation around the beam axis. 
The new alignment procedure also considers the two remaining rotational angles of the sensor, consequently improving the telescope resolution to \SI{5}{\micro\m}. 
For each AC-LGAD spatial and temporal resolution measurement presented in this paper, the contributions from the tracking telescope (\SI{5}{\micro\m}) and MCP--PMT time reference (\SI{10}{\ps}) have been subtracted in quadrature. 
Since the sensors considered have resolutions greater than \SI{15}{\micro\m} and \SI{30}{\ps}, the reference contributions typically have a very small impact.

\section{Signal properties}\label{sec:properties}

\begin{figure}[htp]
    \centering
    \includegraphics[width=0.49\textwidth]{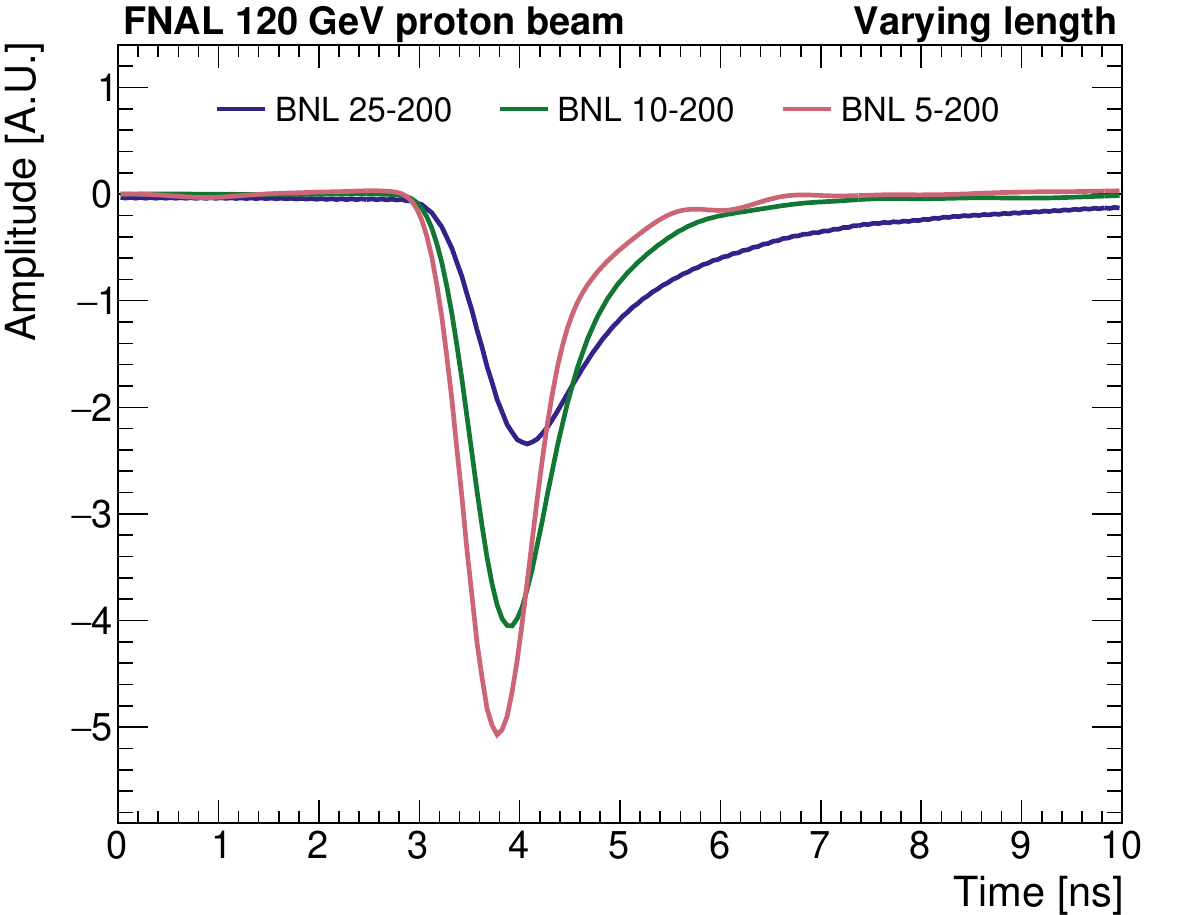}
    \includegraphics[width=0.49\textwidth]{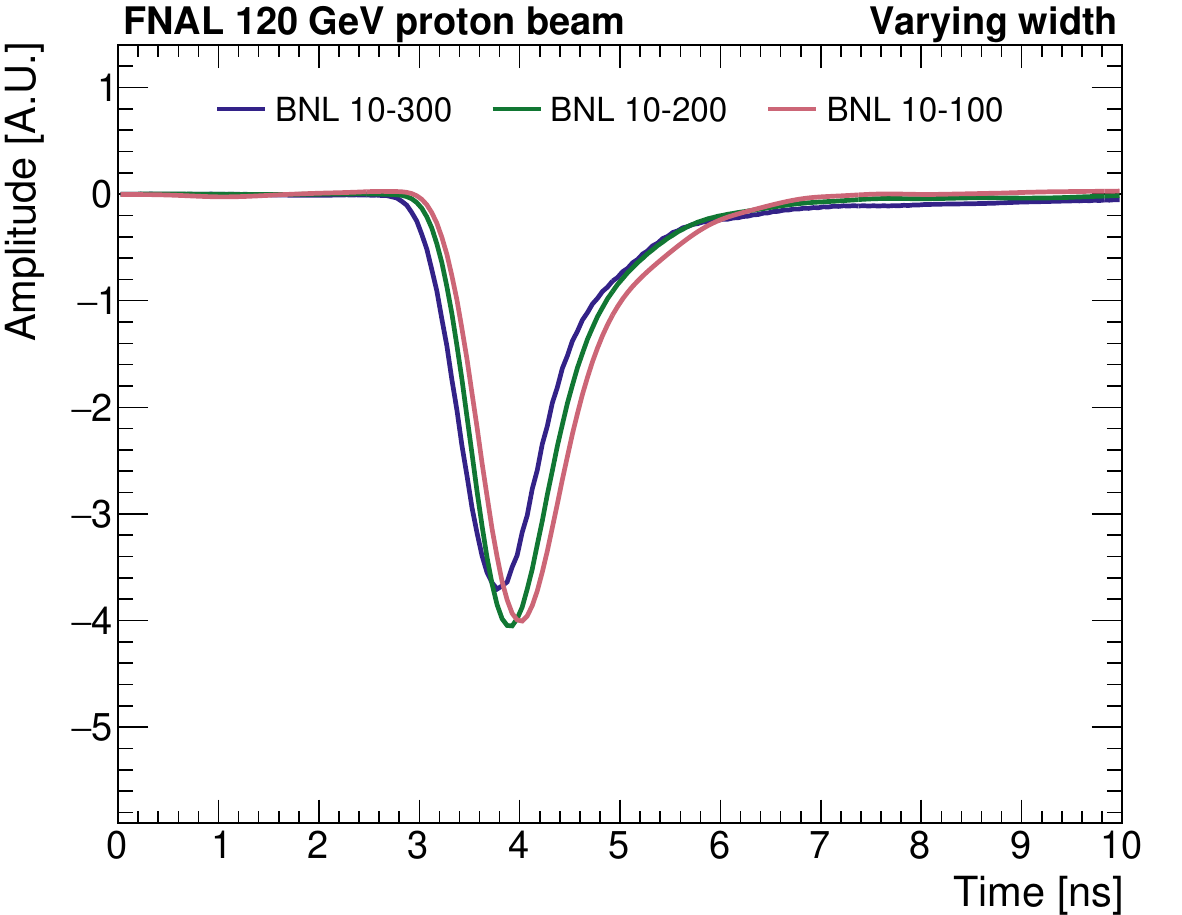}
    \caption{Average waveforms, normalized by area, from sensors with varying lengths (left) and varying metal widths (right).}
    \label{fig:Waveform_DiffLength}
\end{figure}

As described in Section~\ref{sec:setup}, the waveforms from up to seven channels of each AC-LGAD are recorded by the oscilloscope. 
The waveforms are analyzed event-by-event to extract the amplitude, risetime, slew rate, and other relevant signal properties. 
The averaged pulse shapes from the three sensor length variants can be seen in Figure~\ref{fig:Waveform_DiffLength}, normalized by their integrals. 
The timing performance is optimal for shorter sensors as the risetime tends to increase with the length of the sensor.

Events considered in the analysis are required to have a high-quality track and an MCP-PMT timestamp, ensuring reliable references for the proton position and time of arrival. High-quality tracks are required to have hits from all 14 layers of the telescope, reduced $\chi^2 <  3.0$, and slopes less than $10^{-4}$ with respect to the beam axis. Furthermore, only events with a track pointing through the active area of the AC-LGAD are considered. These events form the denominator for all efficiency measurements.
A minimum amplitude is required to identify which channels have signals well-separated from noise. 
The channels used in the multi-strip reconstruction methods, described in the following sections, must have signal pulses with amplitude greater than \SI{15}{\mV}.
To avoid partially contained clusters at the edge of the active sensor area, we only consider events for which the largest amplitude signal is found in a channel that is not at the edge of the sensor. 
The AC-LGADs considered in this paper typically have signals in one or two neighboring strips in each event.

\subsection{Gain uniformity}\label{subsec:uniformity}

\begin{figure}[htp]
    \centering
    \includegraphics[width=0.49\textwidth]{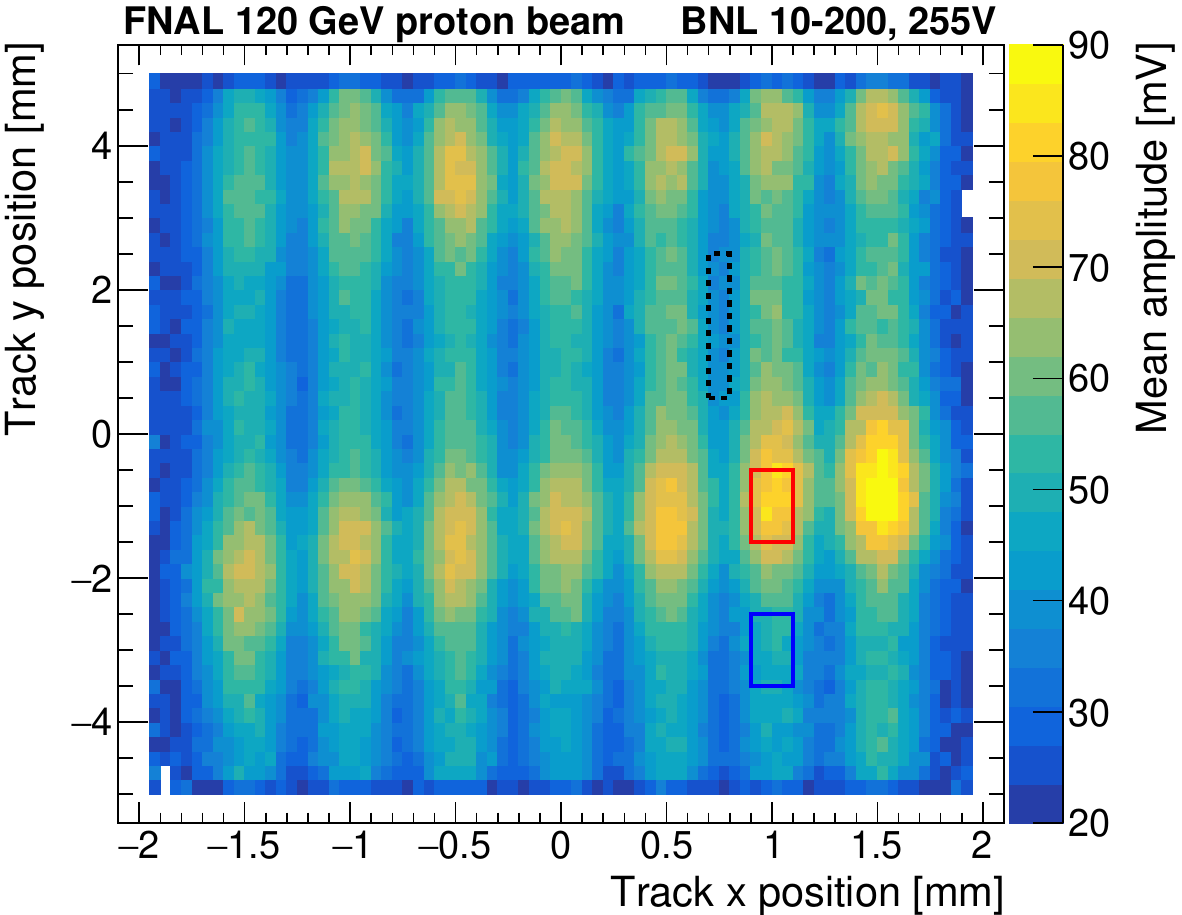}
    \includegraphics[width=0.49\textwidth]{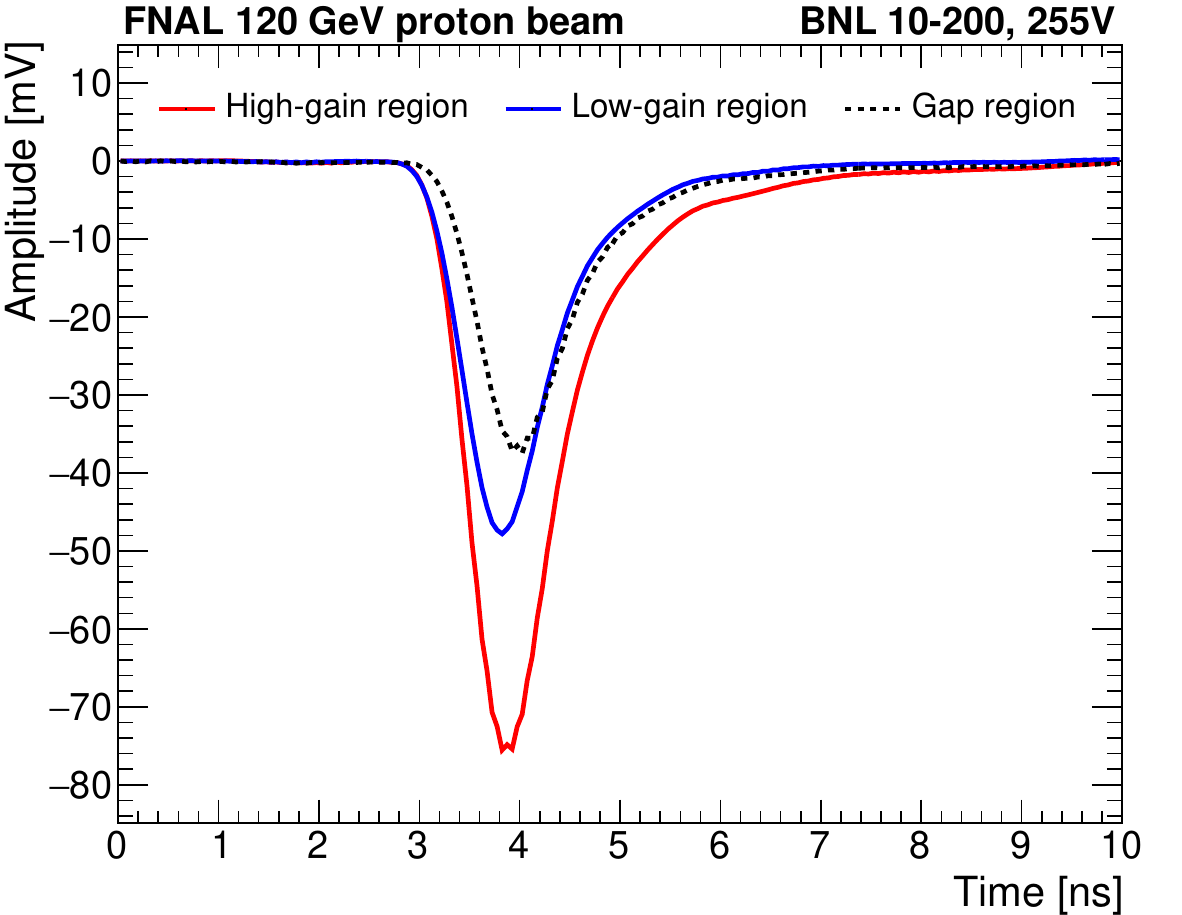}
    \hspace{0.1cm}
    \caption{Signal mean amplitude as a function of the telescope track $x$ and $y$ position (left) for the BNL 10-200 sensor. The different colored boxes represent the high-gain (red), low-gain (blue), and gap (black dotted) regions. The averaged waveform (right) for the three different regions.
    }
    \label{fig:AmplitudeROI_EIC1cm200um}
\end{figure}

\begin{figure}[htp]
    \centering
    \includegraphics[width=0.49\textwidth]{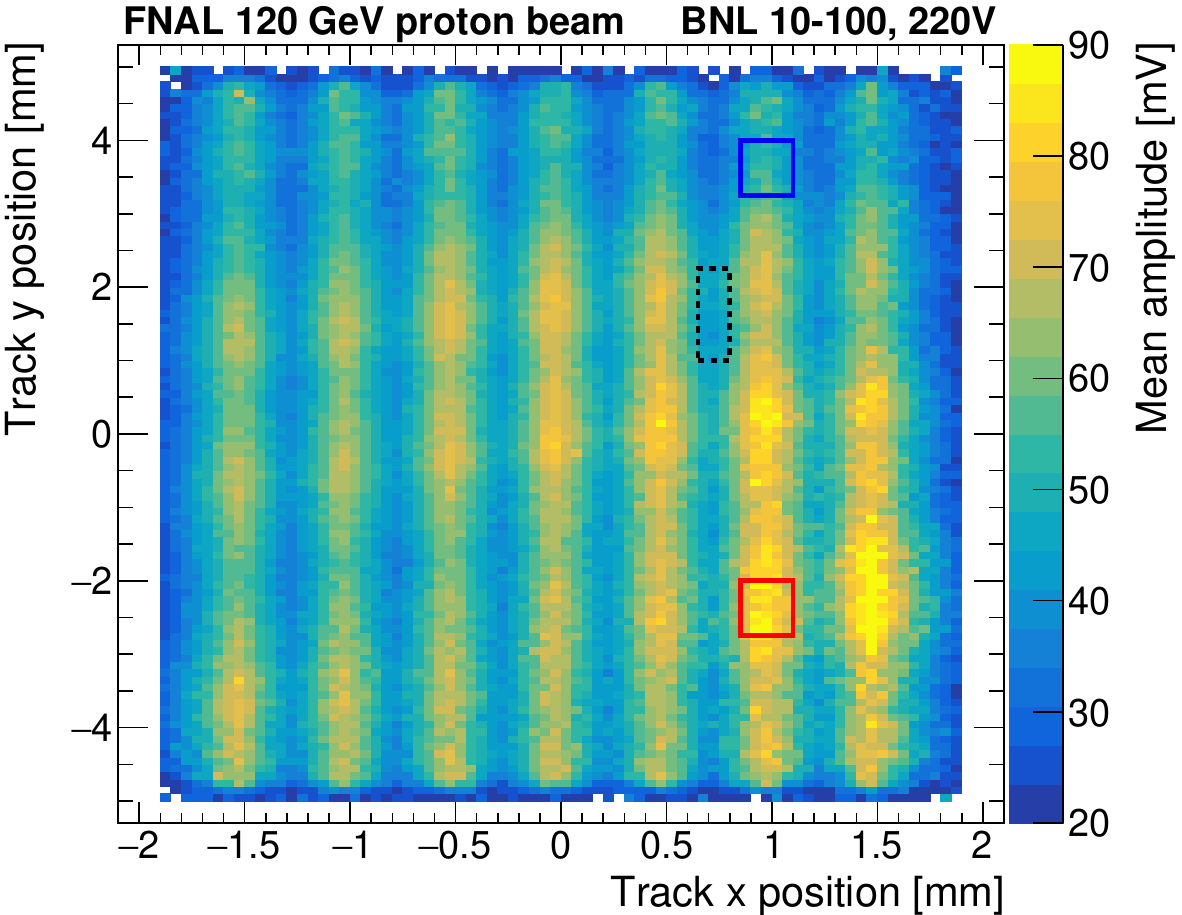}
    \includegraphics[width=0.49\textwidth]{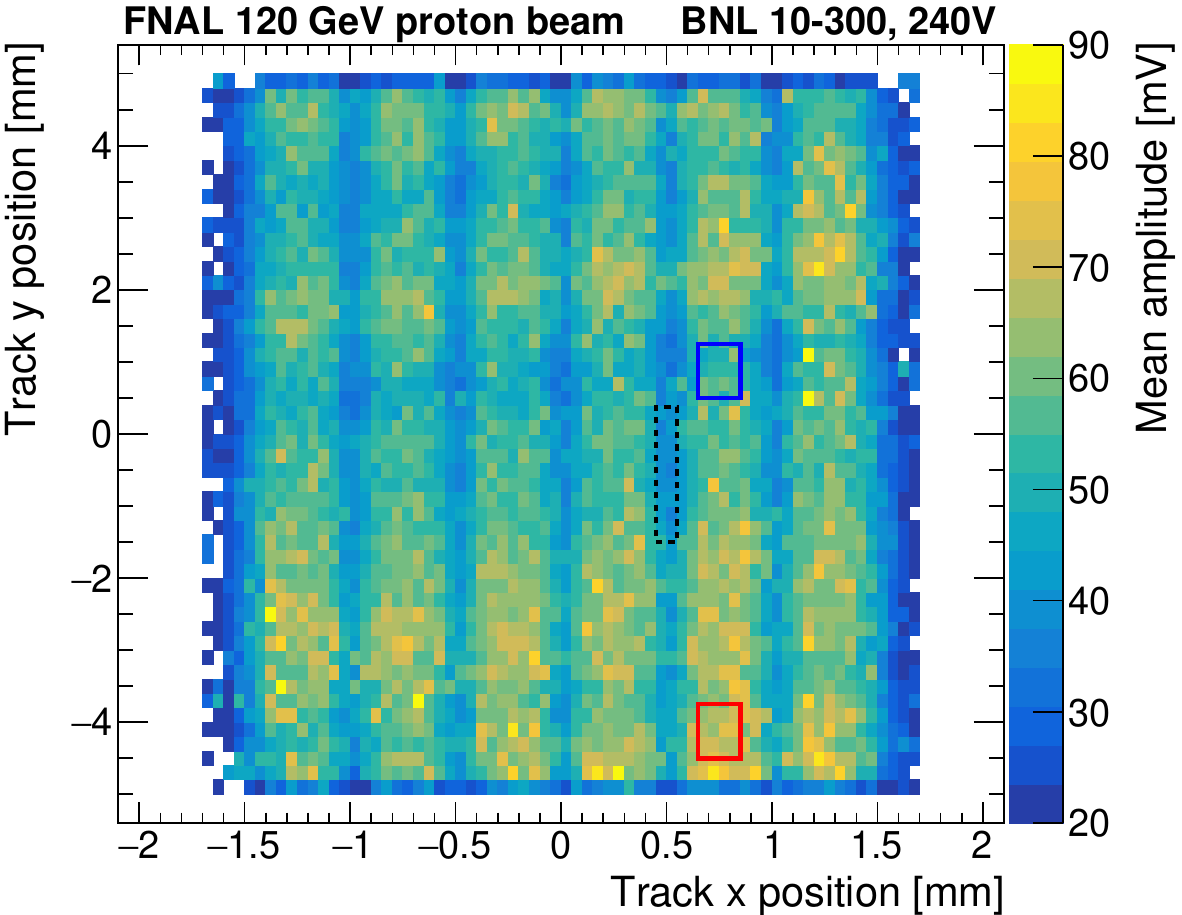}
    \hspace{0.1cm}
    \caption{Signal mean amplitude as a function of the telescope track $x$ and $y$ position for the BNL 10-100 (left) and BNL 10-300 (right) sensors. The different colored boxes represent the high-gain (red), low-gain (blue), and gap (black dotted) regions.}
    \label{fig:AmplitudeROI_EIC1cm100um}
\end{figure}

Gain uniformity across the full area of a sensor is a desirable feature for AC-LGADs.
We observed a degree of non-uniformity in the signal amplitude in all sensors from the current production batch. A characteristic band pattern is observed, as shown in Figure~\ref{fig:AmplitudeROI_EIC1cm200um} and ~\ref{fig:AmplitudeROI_EIC1cm100um}. 
These patterns arise most likely due to non-uniformity in the gain implant. 
The ratio of amplitudes between high and low gain regions is roughly two when the sensors are operated at bias voltage close to the breakdown voltage. 
Because the highest gain regions enter breakdown at lower voltages, they determine the maximum attainable bias voltage for each sensor. 
As result, the lower gain regions remain underbiased, and in some cases, these regions represent the majority of the surface because the high gain structures can be highly localized. 
Many important performance metrics are strongly dependent on gain, and therefore are observed to vary across the surface.

To study the impact of design and geometry choices isolated from the effects of the non-uniformity, we define three different regions within each sensor that characterize the high-gain, low-gain, and gap regions. 
The gap region is the space between adjacent metalized strips, where we also observe differences in signal response. 
Examples of these regions of interest are overlaid in Figure~\ref{fig:AmplitudeROI_EIC1cm200um}, and the corresponding average pulse shapes in each region are also shown. 
We take the performance in the high-gain region as indicative of what would be attained in a uniform sensor, where all regions would reach high gain at similar bias voltage. 
Since uniform large-area LGADs are produced routinely, it is expected that future large-area AC-LGADs productions by BNL or alternative foundries will be able to achieve good uniformity.

In Subsection~\ref{subsec:metrics}, we study a variety of signal shape metrics relevant for the spatial and time resolutions as a function of the strip geometries. 
In order to minimize the effect of the gain non-uniformity, we define the metrics in a gain-independent manner by accounting for or dividing out the dependence on the measured charge.

\subsection{Signal metrics}\label{subsec:metrics}

Achieving good time resolution requires signals with a steep rising edge well-separated from the noise. More precisely, the jitter contribution to the time resolution, representing the impact of the electronic noise, is given by:
\begin{equation}
    \sigma_{\rm{jitter}} = \frac{N}{dV/dt} \sim \frac{t_{\rm{rise}}}{S/N}
     \label{eq:jitter}
\end{equation} 
where $N$ represents the noise amplitude, $dV/dt$ is the slew rate, $t_{\rm{rise}}$ is the risetime, and $S$ is the signal amplitude~\cite{hartmut}. Minimizing the jitter calls for sensors not only with large total charge or gain, but also desirable pulse shapes that result in large $dV/dt$. 

To understand the design elements affecting the jitter, the signal development in AC-LGADs can be factorized into two stages as outlined in~\cite{TORNAGO2021165319}. First, the charge carriers drift within the bulk, are multiplied in the gain region, and induce charges in the n+ surface layer. Then, the induced signal spreads through the resistive layer, and induces signals in the AC-coupled electrodes nearby. The first stage of charge generation and multiplication has little design freedom among LGADs of \SI{50}{\micro\m} thickness with uniform gain. In absence of premature breakdown, these sensors typically deliver the best performance at bias voltage roughly \SIrange[]{10}{15}{\volt} before breakdown, before the onset of additional noise, and where the gain reaches approximately a factor of 20--40. For modest variation of the gain implant doping, the breakdown and operating voltages will shift, but the gain at optimal bias voltage is more or less fixed~\cite{HELLER2021165828}. With saturated drift velocities (bias $> \SI{200}{\volt}$ at room temperature), the signal risetime for the initial charge development can be as fast as \SIrange[]{400}{500}{\pico\second}.

Considering sensors with typical LGAD gain implants and good production uniformity, then, most of the room for design optimization lies in tuning the geometry to best preserve the signal as it is passed to the AC-electrodes in the second stage of signal development. With the survey of varying geometries in this campaign, significant sculpting of the signal due to the electrode design has been observed. In particular, the largest electrodes are associated with significantly slower risetimes, or alternatively, smaller $dV/dt$ for a given input charge. Since many of the jitter contributions in Equation~\ref{eq:jitter} are correlated with the signal size, and the sensors considered in this survey exhibit non-uniform gain, the various contributions cannot be directly compared across sensors to deduce the influence of the electrode geometry. Instead, we rely on several metrics that are deliberately gain-independent, intended to help predict the performance in uniform sensors of the same geometries. The first metric used is the risetime (from 10\% to 90\% of the maximum amplitude), which naturally tends to be scale-invariant. Then, as the amplitude and slew rate ($dV/dt$) are proportional to the gain, we consider the ratios of these quantities to the charge, determined from the integral of each pulse. These ratios serve as gain-independent metrics of the signal shape. The final metric is an estimate of the expected jitter for an example input charge of \SI{10}{\femto\coulomb} (gain of $\sim 15$), found by multiplying $dV/dt / Q$ by \SI{10}{\femto\coulomb} and inserting into Equation~\ref{eq:jitter} with a characteristic noise of \SI{2}{\milli\volt} typical of the 16-channel board amplifiers.

\begin{figure}[htp]
\centering
    \includegraphics[width=0.49\textwidth]{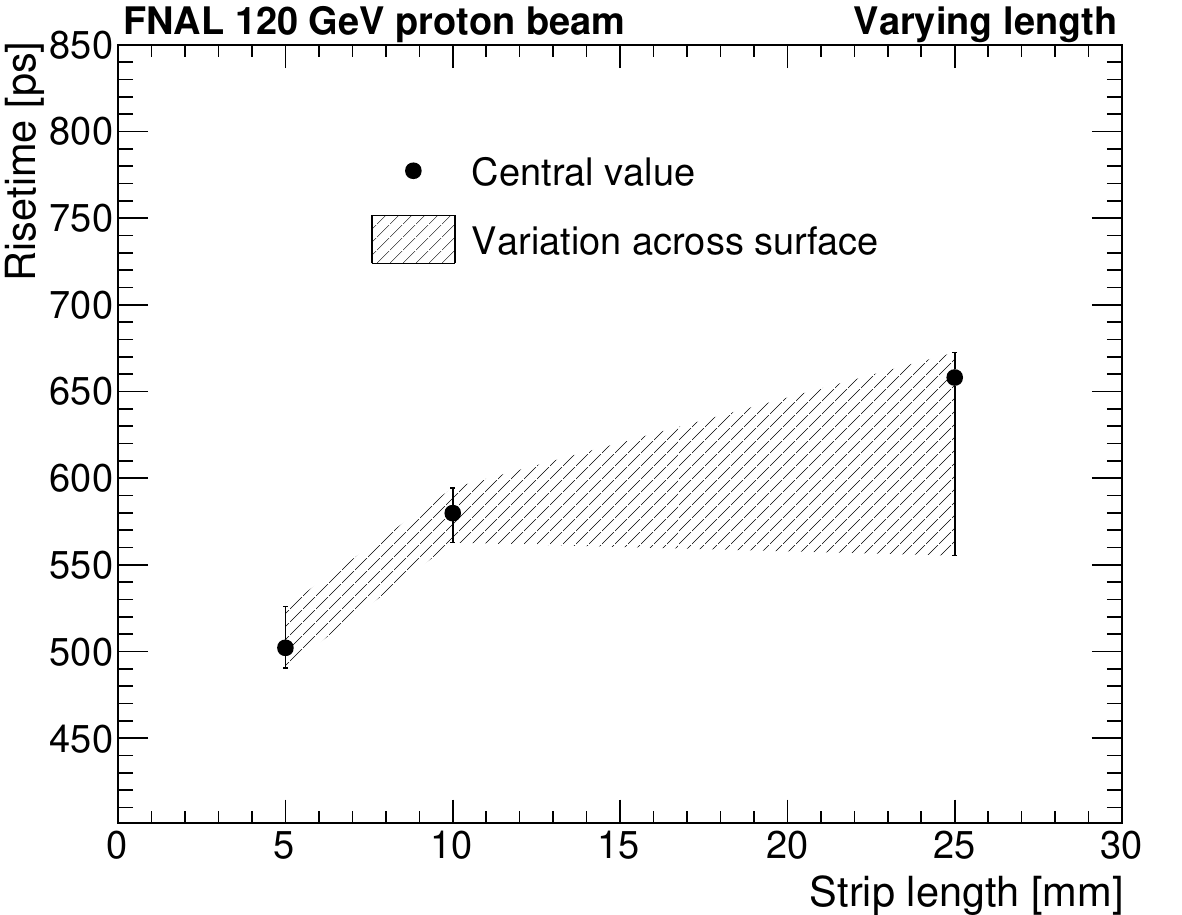}
    \includegraphics[width=0.49\textwidth]{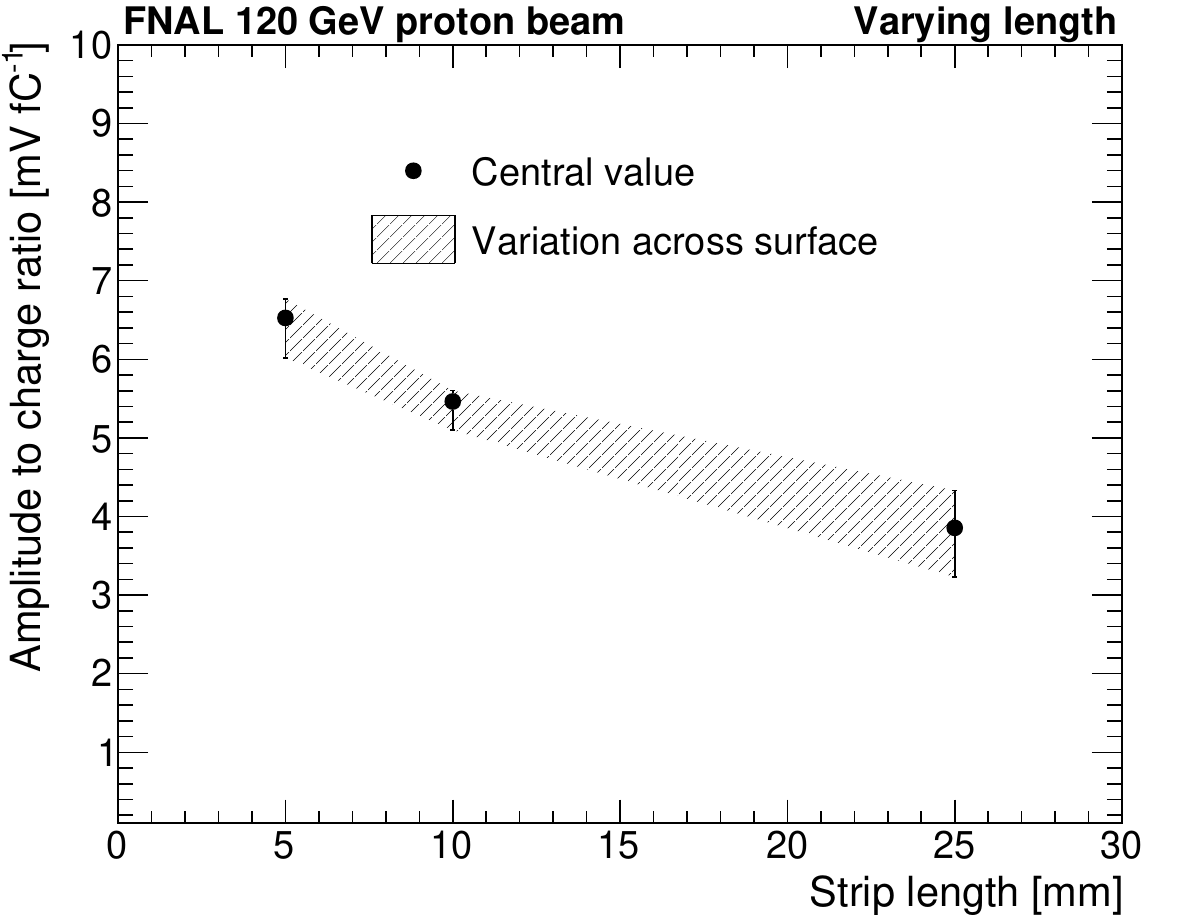}
    
    \includegraphics[width=0.49\textwidth]{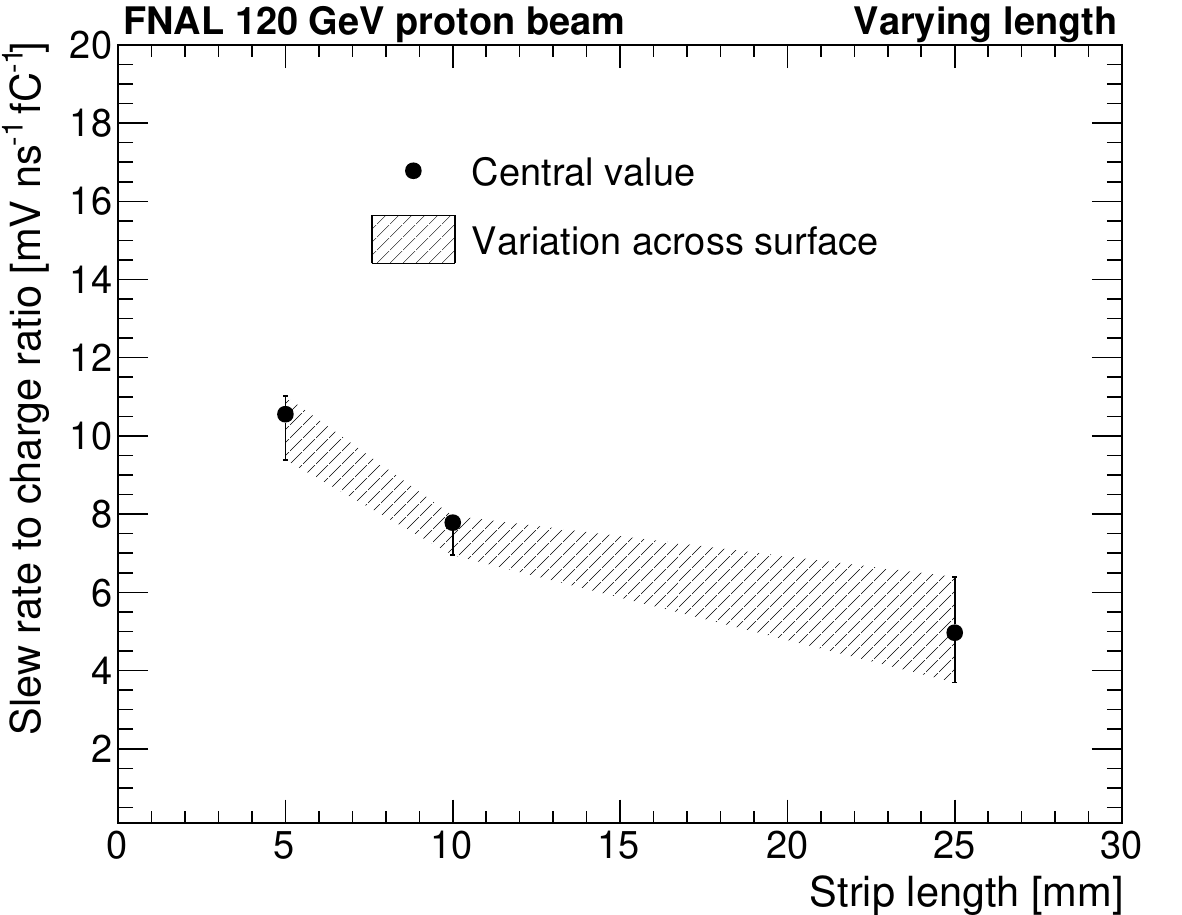}
    \includegraphics[width=0.49\textwidth]{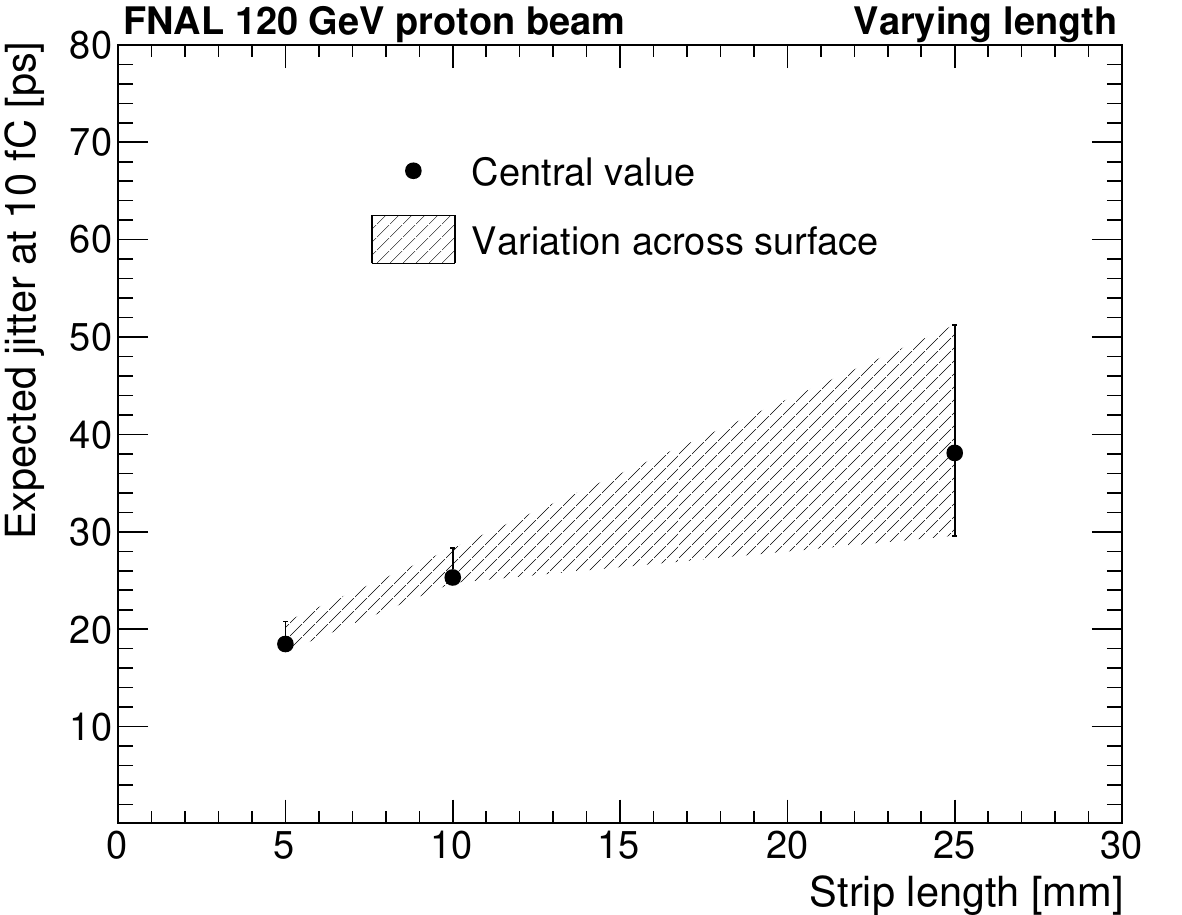}
    \hspace{0.1cm}
    \caption{Risetime (top left), ratio of signal amplitude to charge (top right), ratio of slew rate to charge (bottom left), and expected jitter at \SI{10}{\femto\coulomb} (bottom right) for different length sensors with \SI{200}{\micro\m} electrode width. The uncertainty bands represent variation observed across different regions of each sensor.}
    \label{fig:PulseVar_DiffLength}
\end{figure}

\begin{figure}[htp]
\centering
    \includegraphics[width=0.49\textwidth]{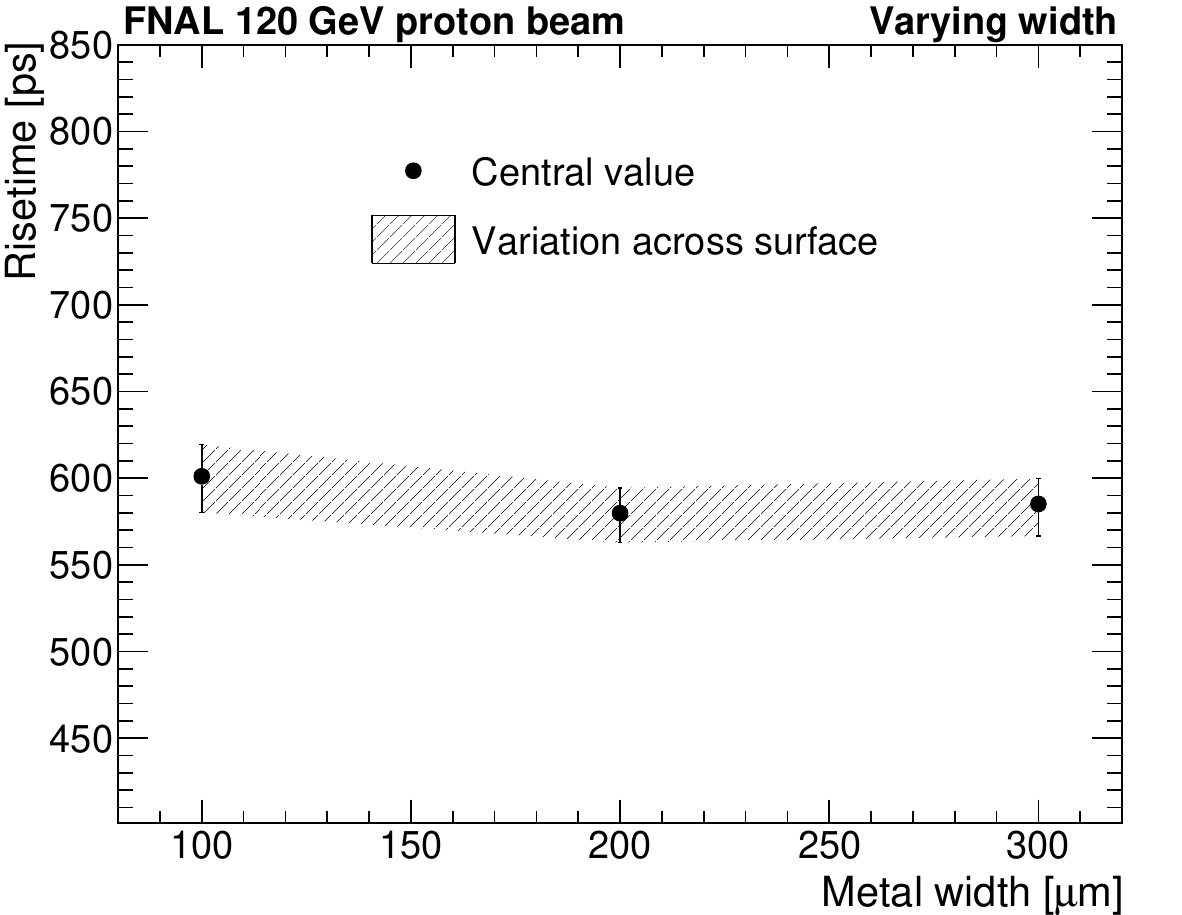}
    \includegraphics[width=0.49\textwidth]{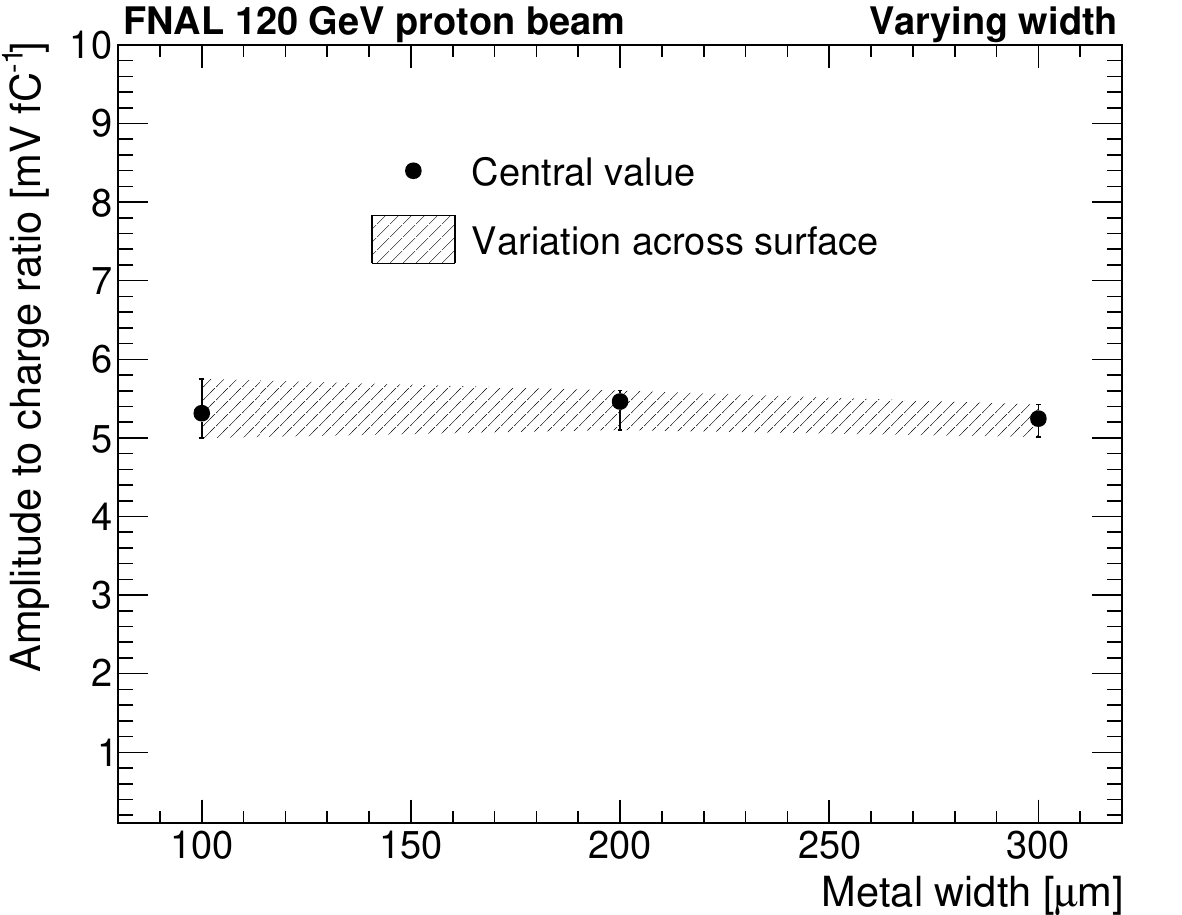}
    
    \includegraphics[width=0.49\textwidth]{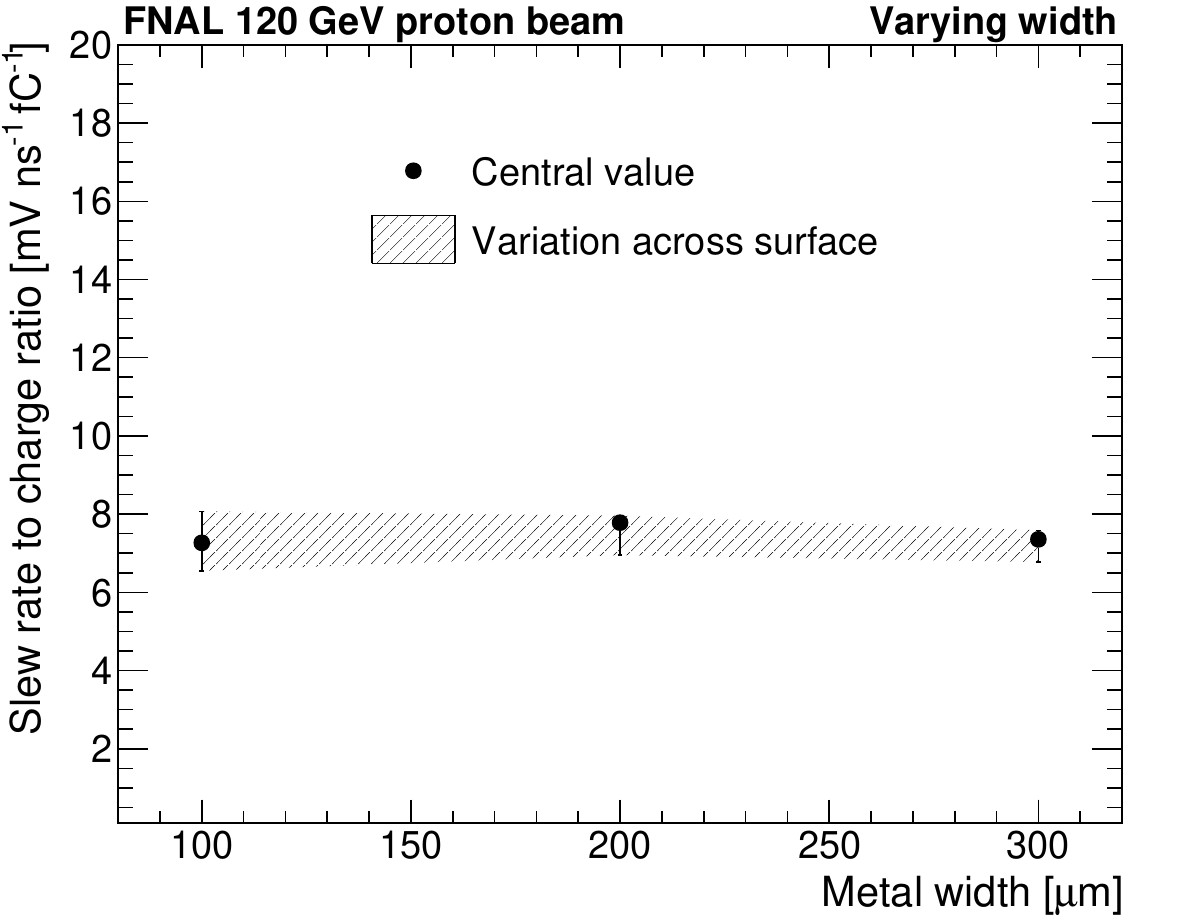}
    \includegraphics[width=0.49\textwidth]{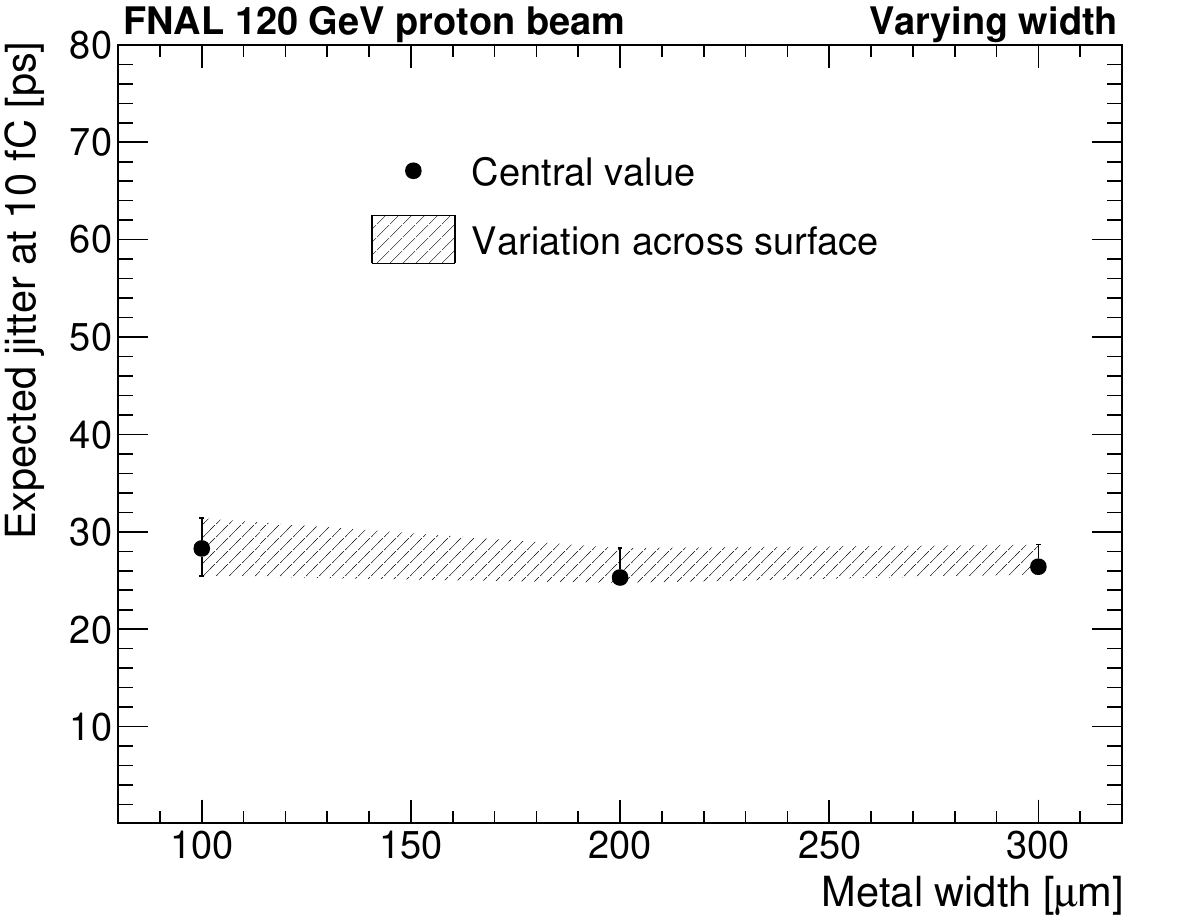}
    \hspace{0.1cm}
    \caption{Risetime (top left), ratio of signal amplitude to charge (top right), ratio of slew rate to charge (bottom left), and expected jitter at \SI{10}{\femto\coulomb} (bottom right) for different metal width sensors with \SI{10}{\mm} long electrode length. The uncertainty bands represent variation observed across different regions of each sensor.}
    \label{fig:PulseVar_DiffWidth}
\end{figure}

These metrics are studied as a function of the strip length with  a constant electrode width of \SI{200}{\micro\m} in Figure~\ref{fig:PulseVar_DiffLength}. The variations observed between the three selected regions of interest described in Section~\ref{subsec:uniformity} form the uncertainty bands. As was visible in the pulse shapes in Figure~\ref{fig:Waveform_DiffLength}, increasing the strip length tends to slow the risetime and yields smaller amplitudes and $dV/dt$ for a given input charge. As result, the expected jitter at a given charge is also larger for the longer strips. Similarly, we consider variations in strip width in Figure~\ref{fig:PulseVar_DiffWidth} at constant \SI{1}{\centi\m} length. In this case, the width does not appear to have a significant impact on the signal metrics. Although all five sensors were operated at slightly different bias voltage, specified in Table~\ref{table:SensorInfo}, it was verified they were all within a range with approximately saturated drift velocities and constant risetimes. As result, the difference in bias voltage does not significantly affect the comparison between sensors.

Two distinct phenomena may contribute to the slowing of the signal observed in the largest strips. First, the slowing may be due to the increased capacitive load on the amplifiers. In this case, simply reducing the electrode width to 50 or \SI{100}{\micro\m} may be enough to compensate for the length in the longest strips and maintain faster signals shapes. Or, the amplifier design could be modified to better compensate for a higher input capacitance. Alternatively, the slowing could be due to an impedance mismatch at the strip-wirebond interface, causing part of the signal to reflect back along the full length of the strip and finally be collected at a small delay, leading to an apparent stretching of the rising edge. To mitigate this effect, the transmission properties at the interface could be improved, using shorter wirebonds as well as redundant bonds for each connection. All of these potential solutions will be studied in upcoming beam test campaigns, ideally leading to both improved understanding of the slowing phenomenon, and improved timing performance in the longest strips. Although the longest strips considered in this campaign do suffer from degraded risetime, it seems possible to design AC-LGAD detector systems with electrode lengths greater than \SI{2}{\centi\m} that maintain the fast signal shapes seen in smaller devices.

\section{Position reconstruction}\label{sec:posreco}

In this section we discuss the reconstruction of the proton impact location, highlighting aspects affected by the large size of these sensors relative to previous measurements. 
By exploiting the signal sharing properties of the AC contacts, the proton hit position (impact parameter) between strips, $x$, can be inferred with resolution much finer than the pitch. 
This reconstruction is presented in Sec.~\ref{subsec:Xposition} and is similar to the previous analysis~\cite{Heller_2022}.

Because adjacent strips are read from alternating ends and the signals propagate with a finite velocity, the time difference between adjacent channels can be used to infer the proton hit position (impact parameter), $y$, along the length of the strip with O(\si{\milli \meter}) precision. This analysis is described in Sec.~\ref{subsec:Yposition}.

\subsection{X position reconstruction}\label{subsec:Xposition}
 
\begin{figure}[htp]
    \centering
    \includegraphics[width=0.49\textwidth]{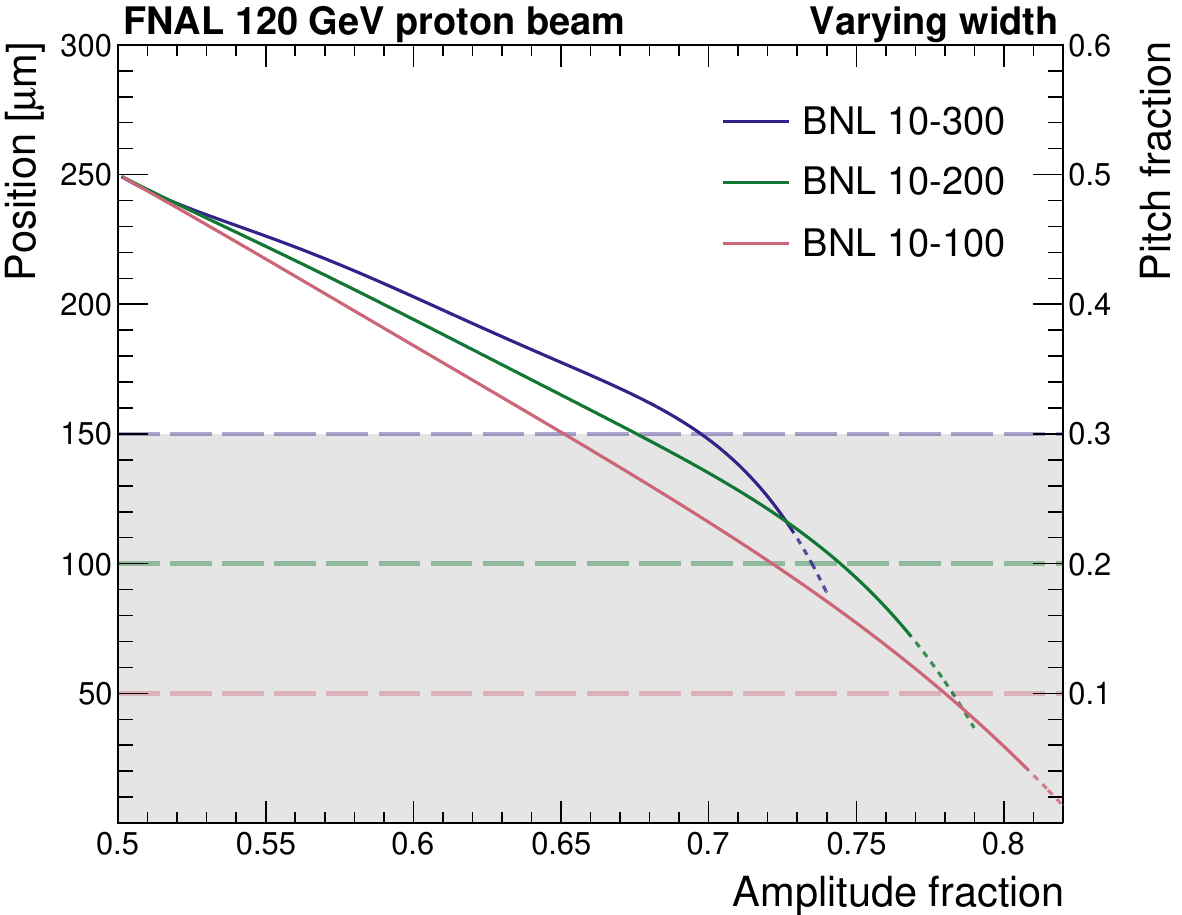}
    \hspace{0.1cm}
    \includegraphics[width=0.49\textwidth]{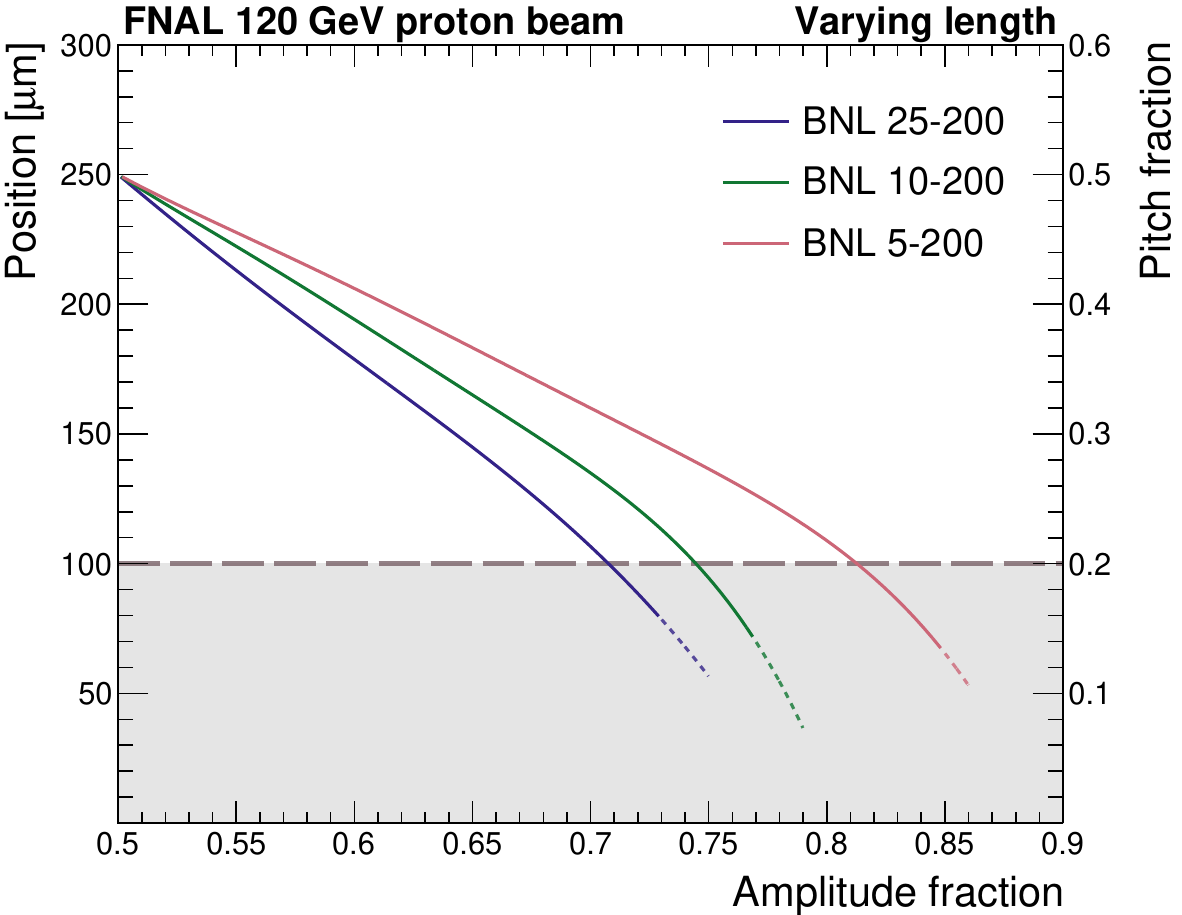}
    \caption{Fit functions, $h(f)$, denoting the proton hit position (impact parameter) as a function of the amplitude fraction, $f$, for varying metal widths (left) and varying strip lengths (right).
    The fits are performed using the full range of the dotted curves, but only the range of the solid curves are used in the reconstruction. The dashed colored lines and grey bands delineate the metalized region of each sensor.}
    \label{fig:PosRecoFit}
\end{figure}

The $x$ position reconstruction takes advantage of the fact that the signal is generally shared between a pair of adjacent strips. 
The reconstruction considers the two highest amplitude strips and interpolates between them to determine the proton hit position (impact parameter). 
To achieve this, we define the amplitude fraction, $f$, as $f=a_1/(a_1+a_2)$, where $a_1$ and $a_2$ are the leading and subleading strip amplitudes. 
Using the tracker reference, we find template histograms of the impact parameter as a function of $f$, and fit them with a polynomial, denoted $h(f)$, unique to each sensor. 
The polynomials $h(f)$ are shown for the different geometries in Figure~\ref{fig:PosRecoFit}. 
The amplitude fraction varies linearly within the gap regions between strips, but saturates upon reaching the boundary of the metalized contact.

In each event the amplitude fraction $f$ is found and the function $h(f)$ is used to determine the impact parameter relative to the leading strip.
This method can be used in events that satisfy two conditions: at least two strips must have an amplitude larger than the minimum threshold; and the amplitude fraction must be less than some value, above which the $f$ no longer varies with the proton impact position and typically occurs close to the metalized strips. 
Events that fail either condition are instead classified as "one strip" events and assigned an $x$ position at the center of the leading strip.

\begin{figure}[htp]
    \centering
    \includegraphics[width=0.49\textwidth]{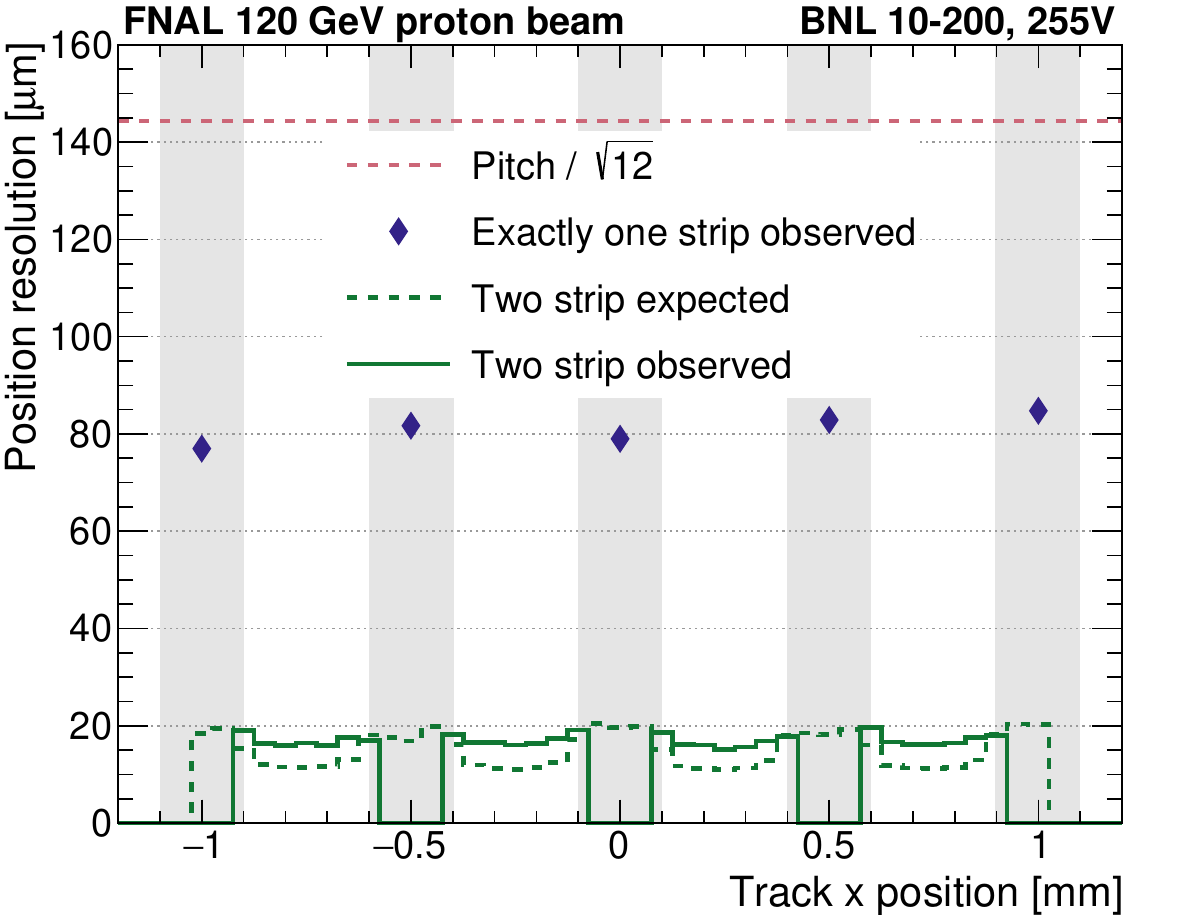}
    \hspace{0.1cm}
    \includegraphics[width=0.49\textwidth]{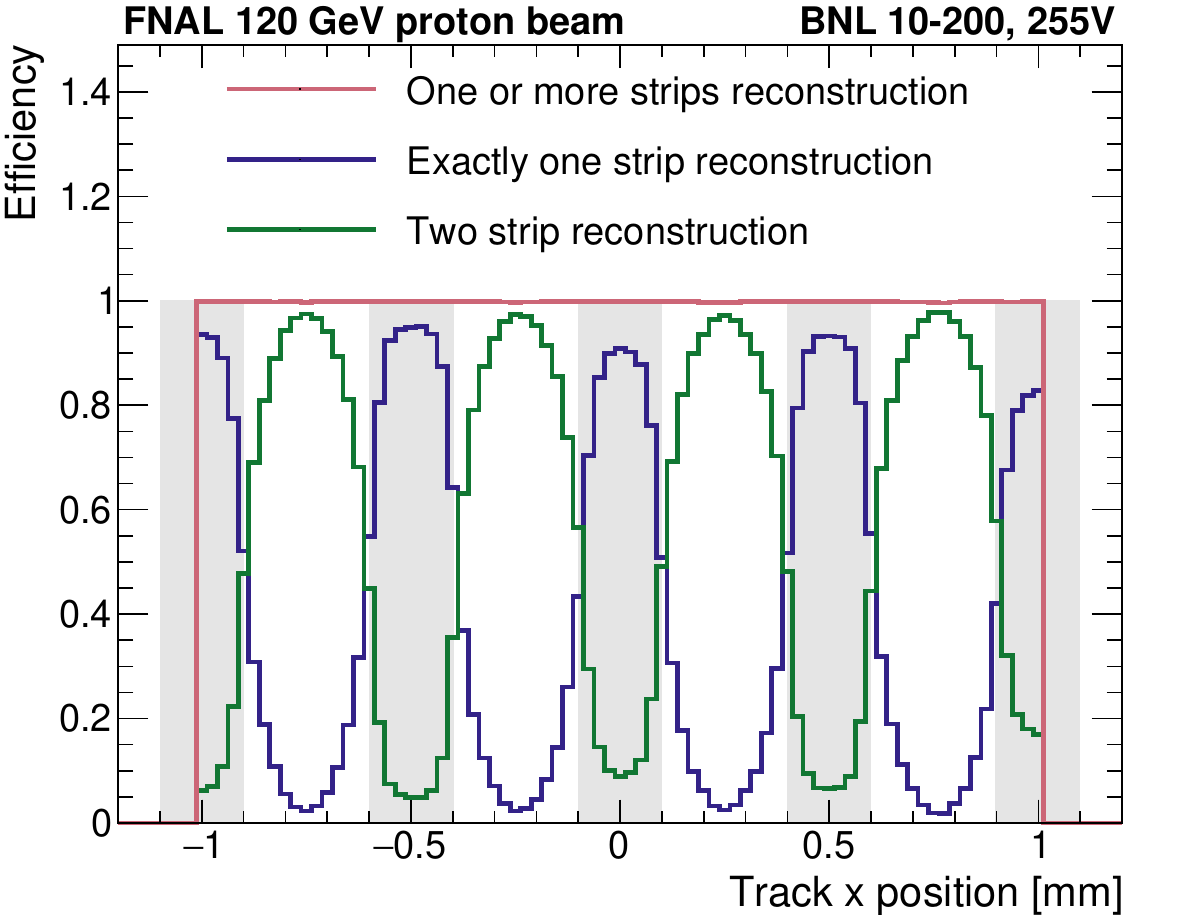}
    \caption{Resolution (left) and efficiency (right) for the $x$ position reconstruction as a function of the track $x$ position for the BNL 10-200 sensor.
    The left figure indicates the resolution that would be achieved with binary readout, pitch $/\sqrt{12}$ (in red), as well as the expected and observed resolution for the two-strip reconstruction (green), and the observed exactly one strip resolution (blue). The efficiency is shown separately for events in the single- and two-strip categories, as well as the union of both categories. The event selection includes hits across the entire sensor surface, including high- and low-gain regions.
    } 
    \label{fig:ResolutionX_200}
\end{figure}

\begin{figure}[htp]
    \centering
    \includegraphics[width=0.49\textwidth]{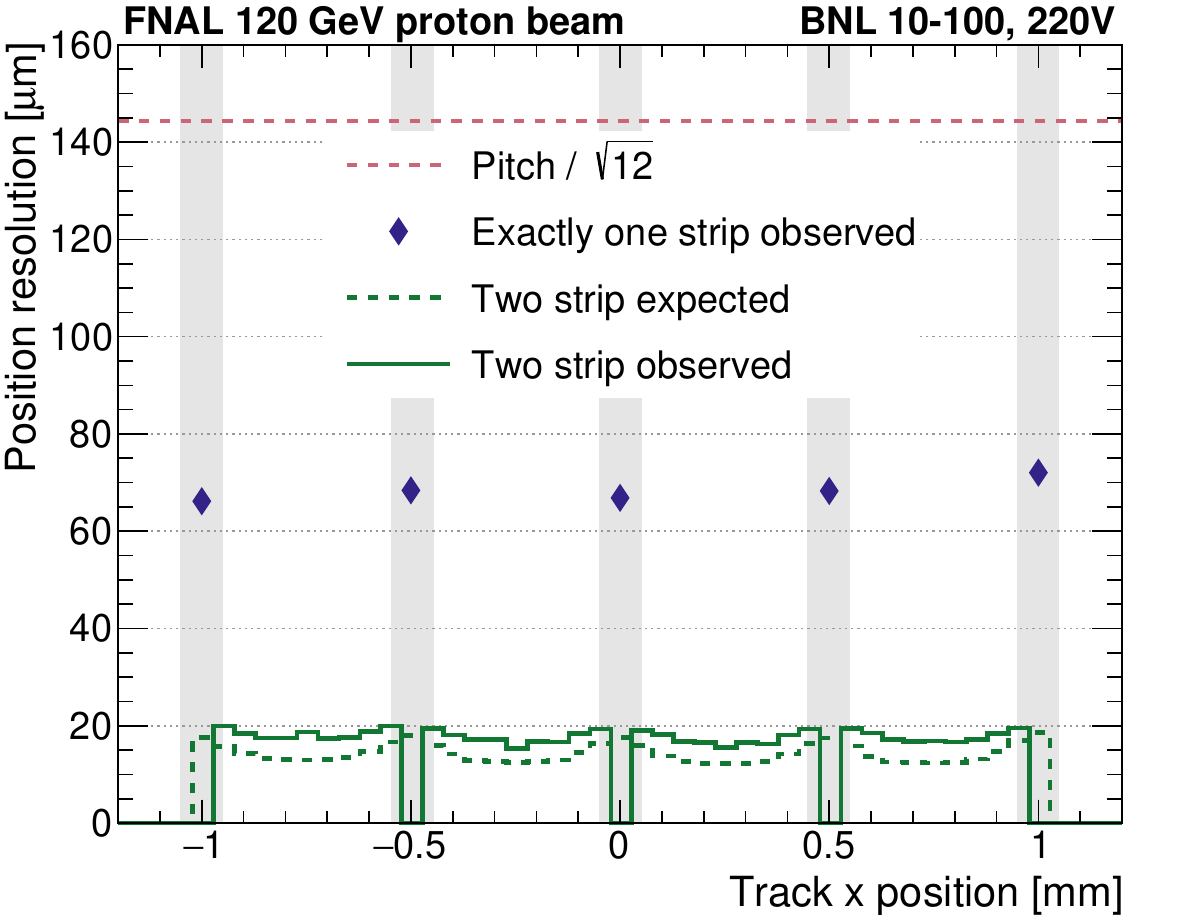}
    \hspace{0.1cm}
    \includegraphics[width=0.49\textwidth]{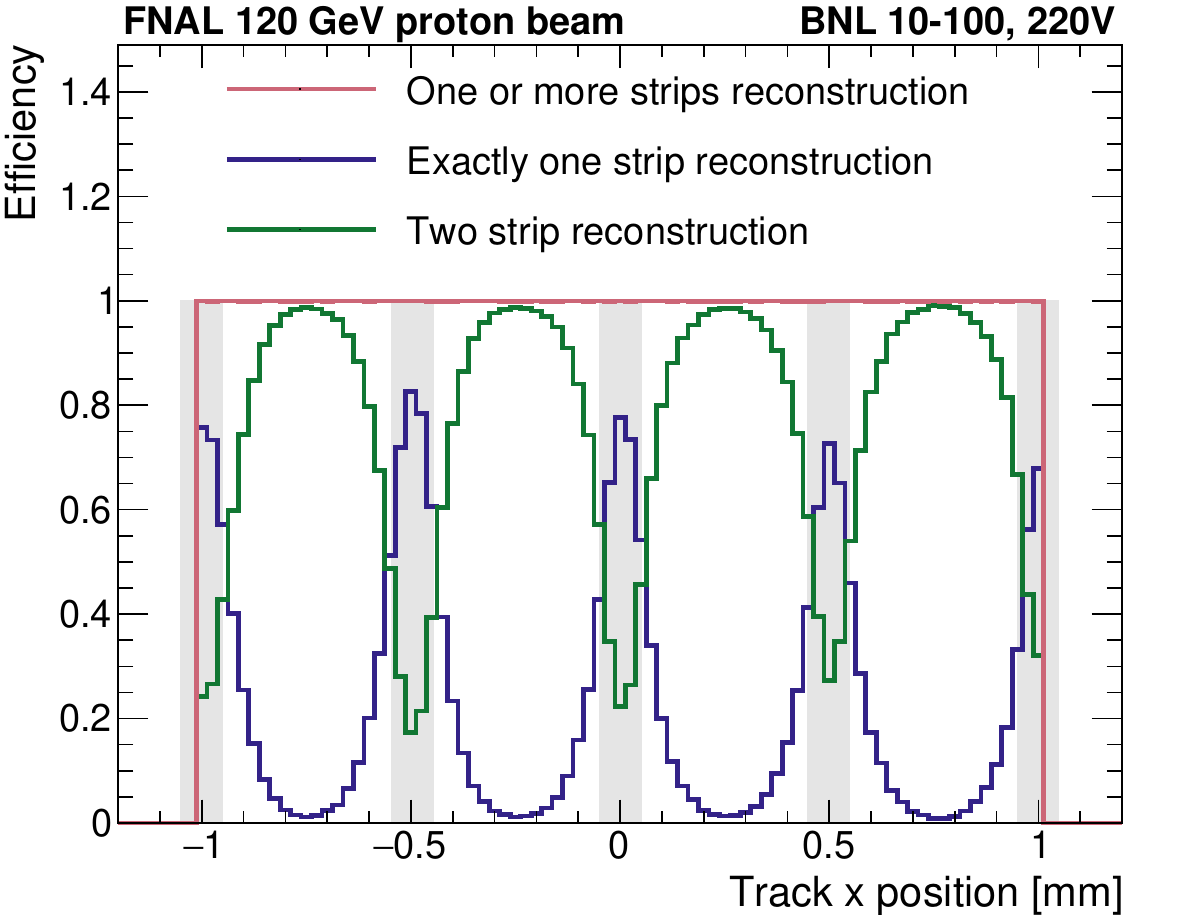}
    \caption{Resolution (left) and efficiency (right) for the $x$ position reconstruction as a function of the track $x$ position for the BNL 10-100 sensor.
    The left figure indicates the resolution that would be achieved with binary readout, pitch $/\sqrt{12}$ (in red), as well as the expected and observed resolution for the two-strip reconstruction (green), and the observed exactly one strip resolution (blue). The efficiency is shown separately for events in the single- and two-strip categories, as well as the union of both categories. The event selection includes hits across the entire sensor surface, including high- and low-gain regions.}
    \label{fig:ResolutionX_100}
\end{figure}

\begin{figure}[htp]
    \centering
    \includegraphics[width=0.49\textwidth]{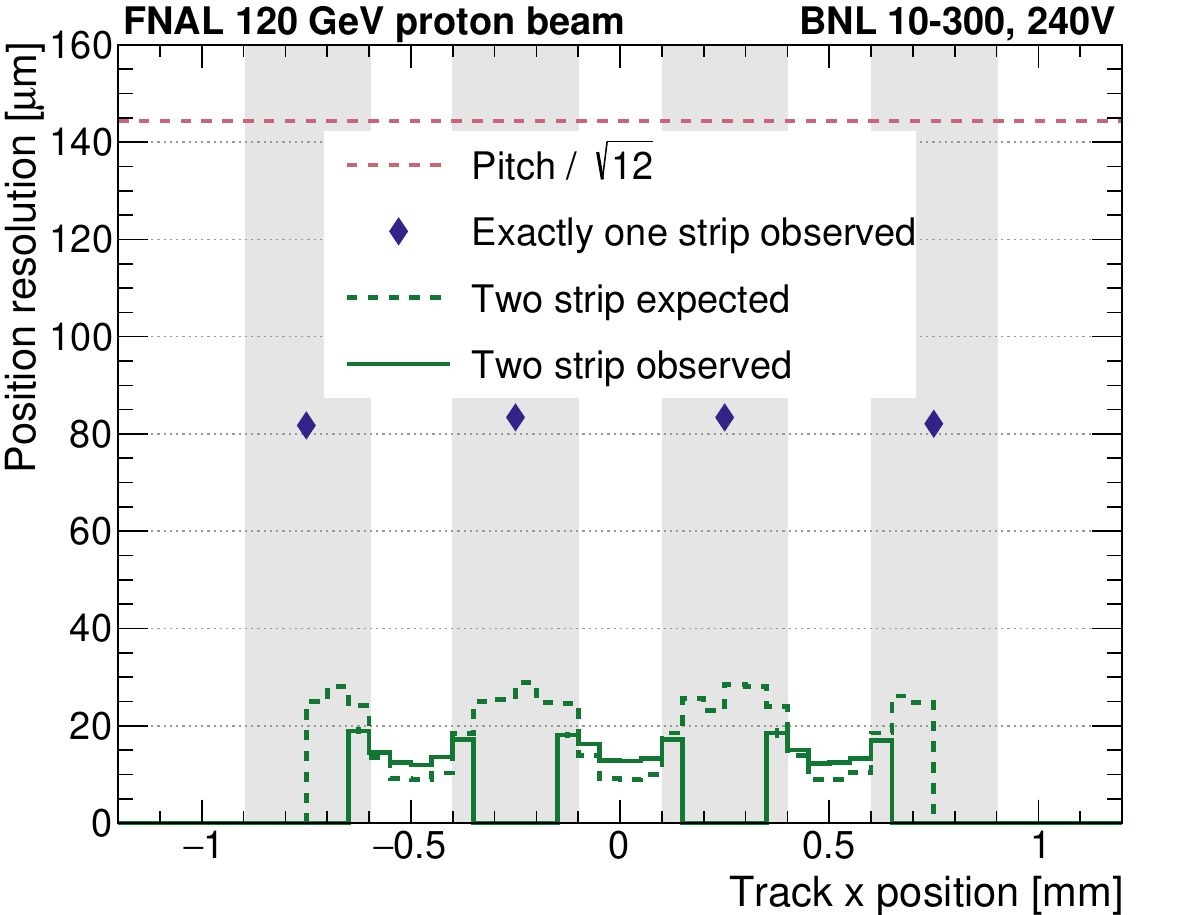}
    \hspace{0.1cm}
    \includegraphics[width=0.49\textwidth]{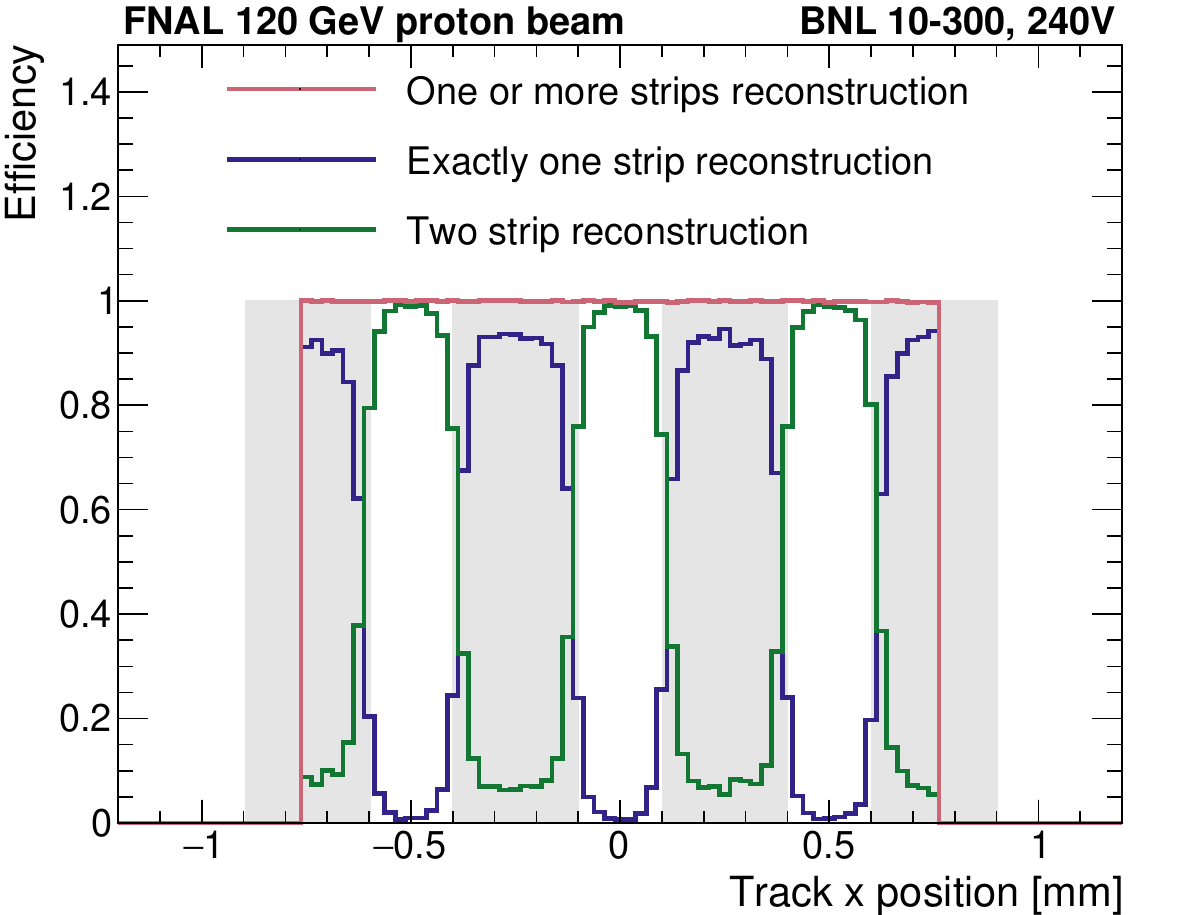}
    \caption{Resolution (left) and efficiency (right) for the $x$ position reconstruction as a function of the track $x$ position for the BNL 10-300 sensor.
    The left figure indicates the resolution that would be achieved with binary readout, pitch $/\sqrt{12}$ (in red), as well as the expected and observed resolution for the two-strip reconstruction (green), and the observed exactly one strip resolution (blue). The efficiency is shown separately for events in the single- and two-strip categories, as well as the union of both categories. The event selection includes hits across the entire sensor surface, including high- and low-gain regions.}
    \label{fig:ResolutionX_300}
\end{figure}

The resulting resolutions and efficiencies for the two reconstruction categories for the three \SI{1}{\cm} sensors are shown in Figures~\ref{fig:ResolutionX_200}, Figure~\ref{fig:ResolutionX_100} and Figure~\ref{fig:ResolutionX_300}, and for all sensors in Table~\ref{table:Summary}, after subtracting a \SI{5}{\micro\meter} contribution from the tracker reference measurement.
The efficiency for a proton to produce a signal in at least one strip reaches 100\%.

When a proton strikes near the center of the gap region, the majority of events fall in the two strip category, and achieves resolution of \SIrange[]{15}{20}{\micro\meter} for all strip widths ranging from 100 to \SI{300}{\micro\m}, significantly smaller than the \SI{500}{\micro\meter} pitch size. 
We derived the expected resolution for the two strip reconstruction by propagating the uncertainty of the noise and signal amplitudes in the two leading strips, $a_1$ and $a_2$, for the signal sharing polynomial $h(f)$:
\begin{equation}
\begin{split}
    \sigma_x^{\rm{expected}} 
    &= \mathrm{P}\left| \frac{dh}{df}\right| \frac{\sqrt{a_1^2+a_2^2}}{(a_1+a_2)^2} N  
    \label{eq:xreco}
\end{split}
\end{equation} 
where P is the sensor pitch and $N$ is the noise amplitude. The expected resolution was evaluated as a function of $x$ and overlaid on Figures~\ref{fig:ResolutionX_200}--\ref{fig:ResolutionX_300}. 
The expected uncertainties are generally quite close to the observed values, but tend to be slightly smaller. 
This difference is attributed to the presence of additional contributions to the observed resolution not captured in Eq.~\ref{eq:xreco}, including the effect of non-uniformity and channel-to-channel variations that are neglected by using a single function $h(f)$ used to characterize the signal sharing. 
Non-uniformity tracker telescope performance could also lead to extra contributions to the observed resolution. 

Signals from protons that strike the metalized strip mostly fall in the single strip category because the majority of the signal is contained within that single strip.
In the absence of a second signal, we must assign its position as the center of the strip, resulting in significantly degraded resolution approaching the limit of the strip width / $\sqrt{12}$. 
The observed resolutions shown in Figures~\ref{fig:ResolutionX_200}--\ref{fig:ResolutionX_300} and Table~\ref{table:Summary} confirm this expectation, though the limit is achieved only for the BNL 10-300 and BNL 5-200 sensors.  
In the other cases, due to the gain non-uniformity, portions of the gap regions with low gain also contribute single strip events. 
These events have larger residuals and spoil the single strip resolution.


The low-gain regions of the prototype sensors compromise the overall performance, since they contribute single-strip events in the gap regions. 
However, sensors with uniform high gain would maintain two-strip efficiency in the gaps and constrain the one-strip events to the metal regions. 
In this case, the single strip resolution would approach the expected limit of metal width / $\sqrt{12}$. 
As result, with sufficiently small metal width (\SIrange[]{50}{100}{\micro\m}), the sensor would yield \SIrange[]{15}{30}{\micro\m} resolution uniformly for both categories across the surface. 
Uniform DC-LGADs of similar size and similar gain implant have already been demonstrated by Hamamatsu Photonics K.K (HPK) and Fondazione Bruno Kessler (FBK) for the CMS and ATLAS timing detector upgrades.
Based on preliminary understanding of the non-uniformity observed in the current production, we expect that improved uniformity and consequently the desired performance will be achieved in the next production from BNL as well.

From Equation~\ref{eq:xreco}, we would expect better two-strip resolution for a smaller derivative of the proton hit position (impact parameter) with respect to signal sharing fraction, $dh/df$. 
Based on the $h(f)$ shown in Figure~\ref{fig:PosRecoFit}, this factor slightly favors the sensors with larger metal width (smaller gap). This trend is observed in the data in Figures~\ref{fig:ResolutionX_200}--\ref{fig:ResolutionX_300} and Table~\ref{table:Summary}, where the \SI{300}{\micro\m} width strips achieve a few micron better resolution than the sensors with smaller metal. This simply reflects the fact that it is easier to interpolate within a smaller gap. However, this effect is relatively small tradeoff given the substantial advantages of increased 2-strip acceptance for sensors with reduced metal width and larger gaps.


\subsection{Y position reconstruction}\label{subsec:Yposition}
Because signals from the strips were read out from alternating ends, the time difference between adjacent channels for events in the two-strip category can be used to reconstruct the $y$ position of the proton hit. 
The $y$ resolution achievable using this method is of order O(\si{\mm}), which is not sufficiently fine to be useful for track reconstruction but is sufficient to correct for position dependent delays that impact the time measurement, discussed in more detail in Section~\ref{sec:timing}. 
This $y$ reconstruction enables both standalone operation without a tracker as well as online timing measurements made before the track reconstruction is available. 
This precision will also assist in pattern recognition by avoiding mismatched hits along long strips.

We can estimate the relationship between the propagation time for signals in two adjacent strips and the $y$ position reconstruction with some simple approximations.
First, we assume the paths the signal takes through the sensor are first approximately aligned with the $x$ direction towards the nearest strips, then along the $y$ direction following the electrode axis.
Second, we assume that the propagation velocity, $v_x$ ($v_y$), along the $x$ ($y$) directions are constant. 
With these two assumptions, an analytic expression for the difference between the $y$ position of the center of the sensor to the proton hit $y$ position ($\Delta y$) can be derived as
\begin{equation}
\begin{split}
    \Delta y 
    &= \frac{v_y}{v_x}\left(\frac{P}{2}-\Delta x\right) + \frac{1}{2}v_y\Delta t
    \label{eq:yreco}
\end{split}
\end{equation} 
where $P$ is the sensor pitch and $\Delta x$ is the $x$ impact parameter. Based on this expression, the expected resolution for the $y$ position estimate is given by:
\begin{equation}
\begin{split}
    \sigma_y^{\rm{expected}} 
    &= \sqrt{\frac{v_y^2}{v_x^2}\sigma^2_x + \frac{1}{2}v_y^2\sigma_t^2}\\
    &\approx \frac{1}{\sqrt{2}}v_y\sigma_t,
    \label{eq:yres}
\end{split}
\end{equation} 
in the limit that $\sigma_x$ is small compared to $\sigma_y$. Note this expression does not depend directly on the sensor length.
With observed velocities $v_y$ on the order of $\SI{50}{\mm/\ns}$, and assuming $\sigma_t = \SI{40}{\ps}$, we expect a $y$ resolution of about \SI{1.4}{\mm}, demonstrating the limit $\sigma_x \ll \sigma_y$.

In these prototypes, the velocities $v_x$ and $v_y$ have nontrivial variation across the surface, which prevents the use of the simple formula in Equation~\ref{eq:yreco}. 
Instead, we derived an empirical lookup table relating the mean $y$ position to the amplitude fraction, $f$, (a proxy for $\Delta x$), and $\Delta t$, which is used to reconstruct the proton $y$ position in each event. 
This procedure yields a resolution governed approximately by Equation~\ref{eq:yres}.

\begin{figure}[htp]
    \centering
    \includegraphics[width=0.49\textwidth]{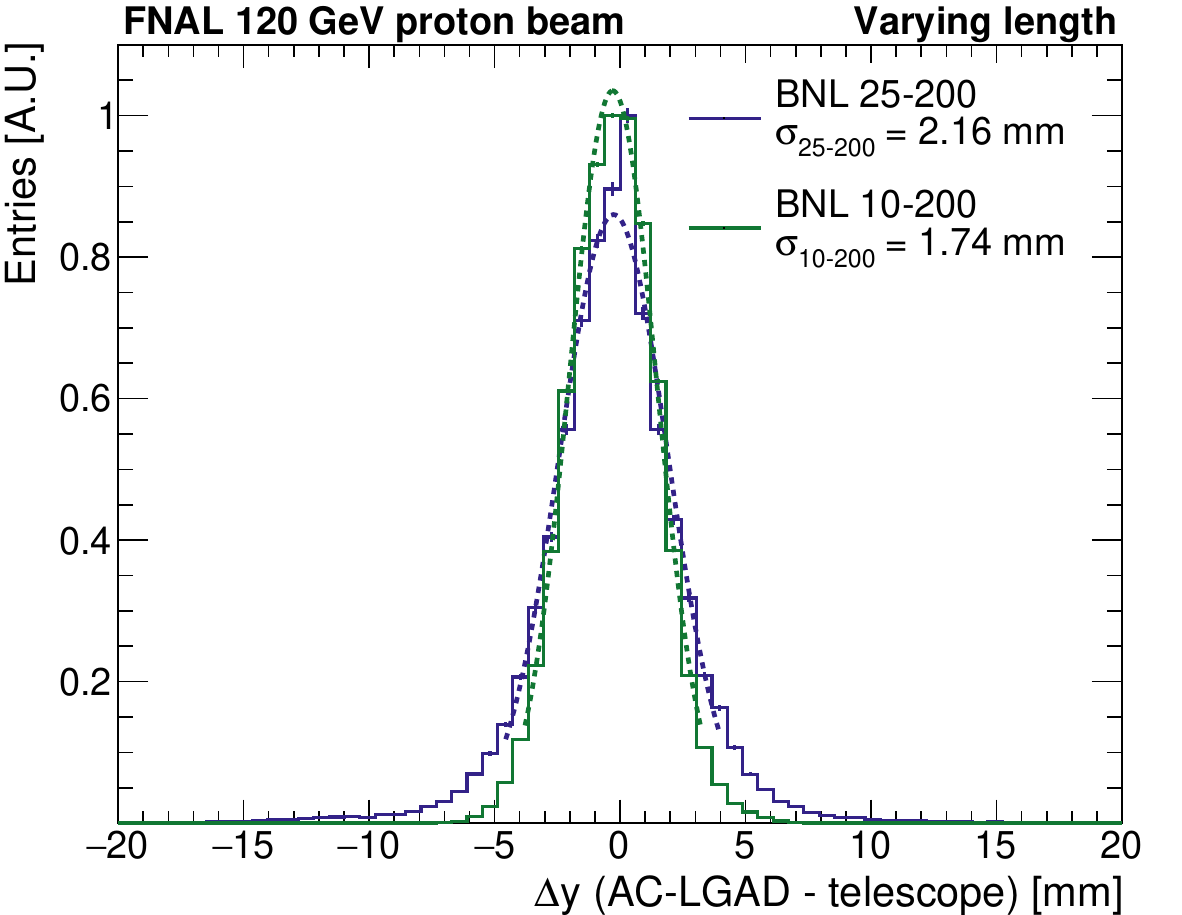}
    \hspace{0.1cm}
    \caption{Residuals of the $y$ position reconstruction for the two longer sensors, BNL 10-200 and 25-200, with Gaussian fit to extract the longitudinal resolution.}
    \label{fig:DeltaY_DiffLength}
\end{figure}

The measured $y$ position resolutions for the BNL 10-200 and BNL 25-200 sensors are shown in Figure~\ref{fig:DeltaY_DiffLength}. The BNL 10-200 exhibits slightly better performance, close to the expected value of \SI{1.4}{\mm}. 
The longer BNL 25-200 sensor has coarser time and $x$ position resolution due to its longer pulse shape and gain non-uniformity, which results in slightly worse $y$ position as well. 
Since the $y$ resolution is not strongly dependent on the strip length, the $y$ measurement would become increasingly more useful for devices that have length larger than several \si{\cm}, able to achieve resolution several times smaller than the length / $\sqrt{12}$.

\section{Time measurements}\label{sec:timing}

\begin{figure}[htp]
    \centering
    \includegraphics[width=0.49\textwidth]{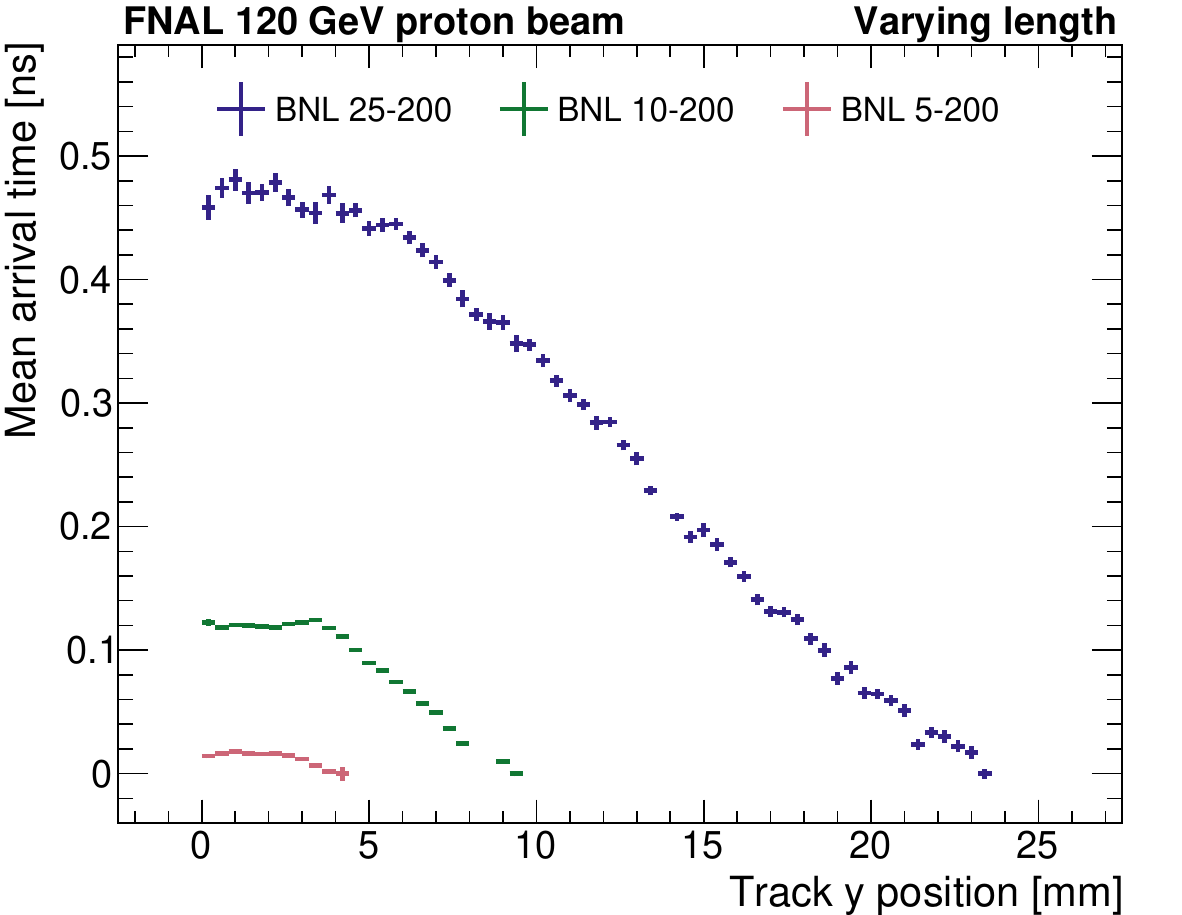}
    \includegraphics[width=0.49\textwidth]{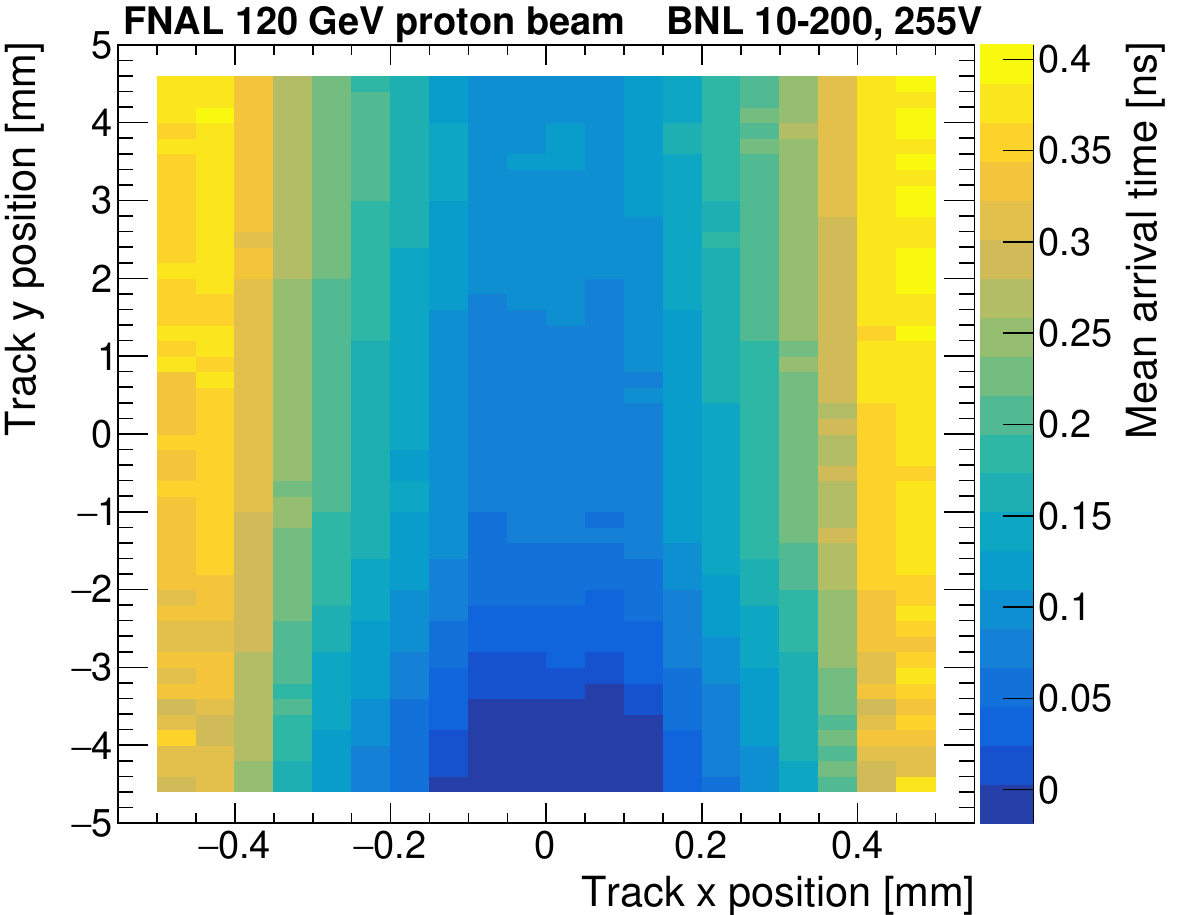}
    \hspace{0.1cm}
    \caption{Variation in mean signal arrival time as a function of the proton impact location along the strip for three length sensors (left). Map of the arrival time for a single channel from  the BNL 10-200 sensor (right). The mean arrival time is defined to be zero for hits near the wirebond at the end of the strip.}
    \label{fig:TimeDealy_DiffLength}
\end{figure}

The increased size of these sensors, particularly along the strip length,  introduces a position-dependent time delay, shown in Figure~\ref{fig:TimeDealy_DiffLength}.
If left uncorrected, this position-dependent time delay significantly degrades the resolution of the time measurement.
Given the centimeter-scale length, the variation in arrival time can be as large as \SI{400}{\ps} and \SI{1}{\ns} for the \SI{10}{\mm} and \SI{25}{\mm} sensors, respectively. 

We demonstrate two strategies to correct for the position-dependent time delay.
The first strategy is to use the external tracker to determine the proton hit position.
The second strategy is to use the position reconstruction procedure outlined in Section~\ref{sec:posreco} to determine the proton hit position. 
In general, the delay correction does not require extremely fine position resolution to remove this contribution to the time resolution, and both strategies perform adequately. 

\subsection{Time delay corrections}
To correct the time delay, we first generate maps of the mean arrival time for each channel on every sensor as a function of proton $x$ and $y$, as in Figure~\ref{fig:TimeDealy_DiffLength} (right). 
Then, in each event, we use the map and the reconstructed $x$ and $y$ position of the proton hit to determine the appropriate correction for the arrival time delay.
The proton $x$ and $y$ position can be found using either the external tracking information, or the AC-LGAD position reconstructions described in Section~\ref{sec:posreco}.

\begin{figure}[htp]
    \centering
    \includegraphics[width=0.49\textwidth]{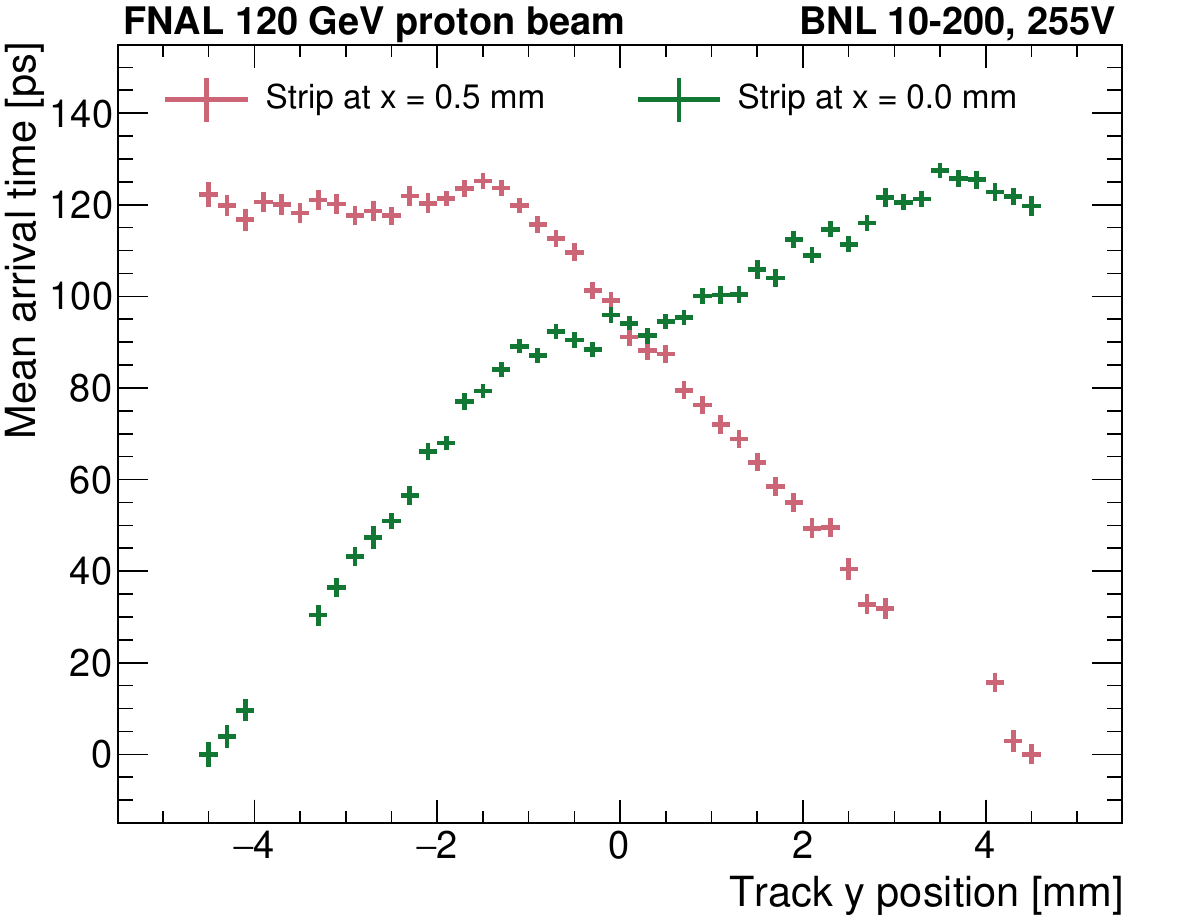}
    \includegraphics[width=0.49\textwidth]{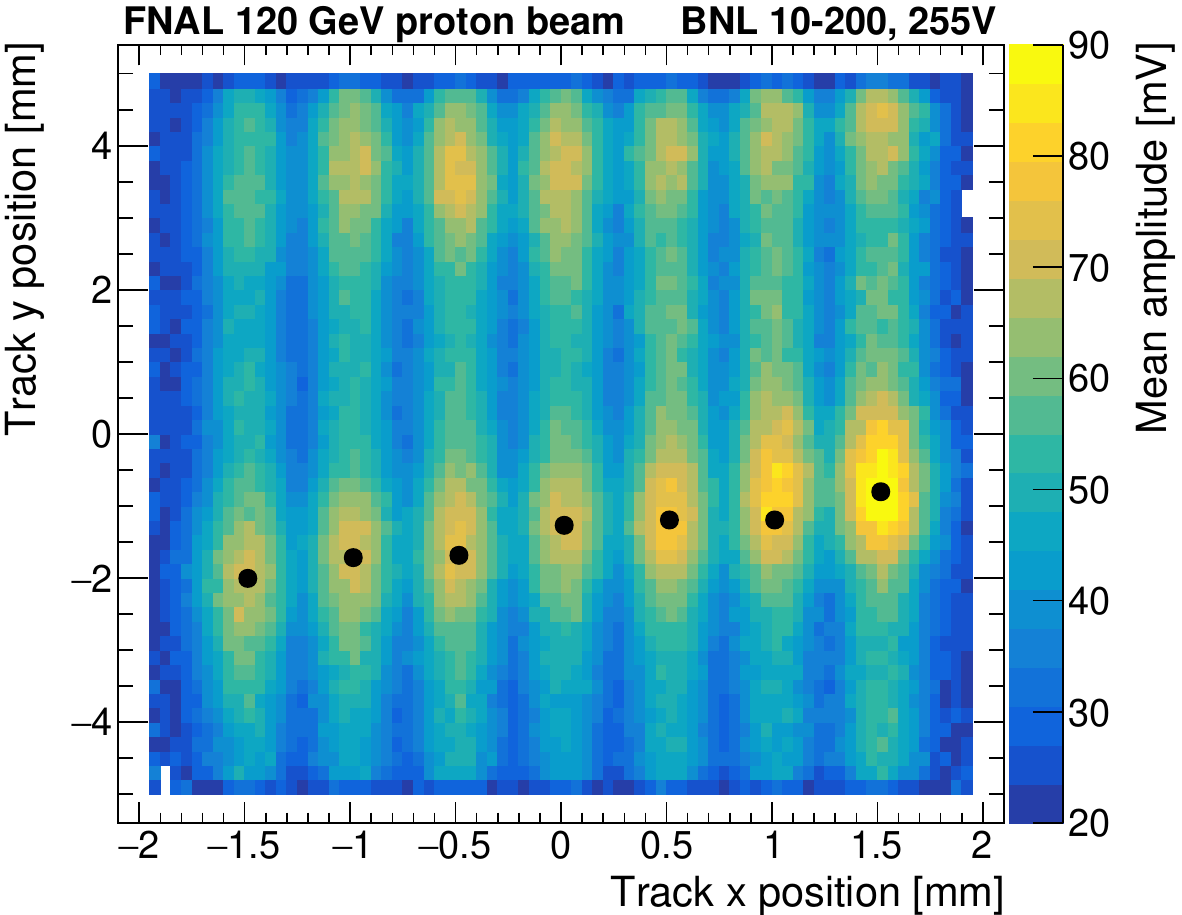}
    \hspace{0.1cm}
    \caption{Mean arrival time as a function of the track $y$ position for two neighboring strips that are read from alternating ends (left). The transition points near $y=-2$~\si{\mm} between regions with different apparent velocities are overlaid as black dots on the map of signal amplitude (right). The transitions between domains occurs at the locations with highest gain, regardless of the readout end.}
    \label{fig:TimeDealy_Transition}
\end{figure}

The form of the delay correction could be further simplified if the delay could be parametrized in terms of a simple function of the $x$ and $y$ positions in each event. 
However, the delays exhibits nontrivial position dependence, as can be seen in the BNL 10-200 sensor in Figure~\ref{fig:TimeDealy_Transition}. 
In general, the sensors exhibit linearly varying delays across large regions, consistent with a constant longitudinal propagation velocity. 
However, frequently these regions are separated into two or more domains with different apparent velocities. 
In some cases, as in BNL 10-200, the inflection points between two domains is strongly correlated with features in the gain non-uniformity. 
In Figure~\ref{fig:TimeDealy_Transition} (right), the delay domain boundary points are overlaid as black points on the map of the signal amplitude variation, and it is clear that they correspond with the locations of peak gain. 
Sensors with higher uniformity tend to have less complex delay structures, but understanding this behavior in detail is beyond the scope of this work. 
For this analysis, we do not use an analytic parametrization of the delays and simply rely on the empirical delay maps as in Figure~\ref{fig:TimeDealy_DiffLength}. We note that within the non-metalized regions between strips, significant variation in arrival time along the $x$ axis is observed, with a delay accumulating with distance from the strip at roughly \SI{700}{\pico \second \per \milli \m}. 
This variation would correspond to an unphysically slow propagation velocity and its origin is not yet fully understood. 

\subsection{Time resolution}

The three main contributions to the AC-LGAD time resolution are the Landau fluctuations in the ionization depth profile, the noise-induced jitter, and the effect of the arrival time delays. 
\begin{equation}
\begin{split}
    \sigma_t^2 &= \sigma_{\rm{Landau}}^2 + \sigma_{\rm{jitter}}^2 + \sigma_{\rm{delay}}^2
\end{split}
\end{equation} 

The Landau term grows with the active thickness of the sensor and sets an approximately \SI{30}{\pico\s} resolution floor for \SI{50}{\micro\m} sensors. 
The jitter term is proportional to the ratio of the noise and the slewrate of the AC-LGAD signal. 
The arrival time variation, left uncorrected, would contribute proportionally to the dimensions of the sensor:
\begin{equation}
    \sigma_{\rm{delay_y}} \sim \frac{L/v_y}{\sqrt{12}}, \quad \sigma_{\rm{delay_x}} \sim \frac{P/v_x}{\sqrt{12}}
\end{equation} 
where $L$ and $P$ are the length and pitch of the sensor. If the delays are instead corrected, the residual contribution to the time resolution due to imprecision in the correction is given by:
\begin{equation}
\begin{split}
    \sigma_{\rm{delay_y}} \sim \frac{\sigma_y}{v_y} , \quad \sigma_{\rm{delay_x}} \sim \frac{\sigma_x}{v_x}.
\end{split}
\end{equation} 
Considering a typical value of $v_y \sim \SI{50}{\mm/\ns}$, the $y$ delay contribution can be brought to a negligible \SI{10}{\pico\s} with $\sigma_y \sim \SI{500}{\micro\m}$. 
These resolutions are easily achievable within a tracking system, and are nearly achieved in standalone operation as well.
As result, the time resolution even with large sensors remains dominated by the Landau fluctuations and in some cases, the jitter. 
Sensors with uniform, high gain would have improved $\sigma_{\rm{jitter}}$, $\sigma_y$, and $\sigma_x$, and should achieve resolutions at the Landau floor of \SI{30}{\pico\second}.

\subsection{Time resolution results}
Signal sharing between the strips can also be utilized to improve the time measurement of the AC-LGAD sensor, as shown in previous results~\cite{Heller_2022}. 
The simplest approach we use to reconstruct the time for a given hit is to use the timestamp measured in the channel with the largest signal amplitude, using a constant fraction discriminator (CFD) algorithm at 50\% level to the leading edge of the signal. 
However, this approach can lead to non-uniform time resolution across the sensor as the single-channel signal amplitudes are smaller in the gaps between two channels. 

An improved method calculates timestamps according to the amplitude squared weighted average of multiple channels as follows:
\begin{equation}
\begin{split}
    t_{\rm{reco}} 
    &= \frac{a_1^2t_1 + a_2^2t_2}{a_1^2 + a_2^2}
    \label{eq:muti-channel_timestamp}
\end{split}
\end{equation} 
where $a_{1}$ and $a_{2}$ are the amplitude of the leading and subleading channels and $t_{1}$ and $t_{2}$ are the time measurements of leading and subleading channels.

\begin{figure}[htp]
    \centering
    \includegraphics[width=0.49\textwidth]{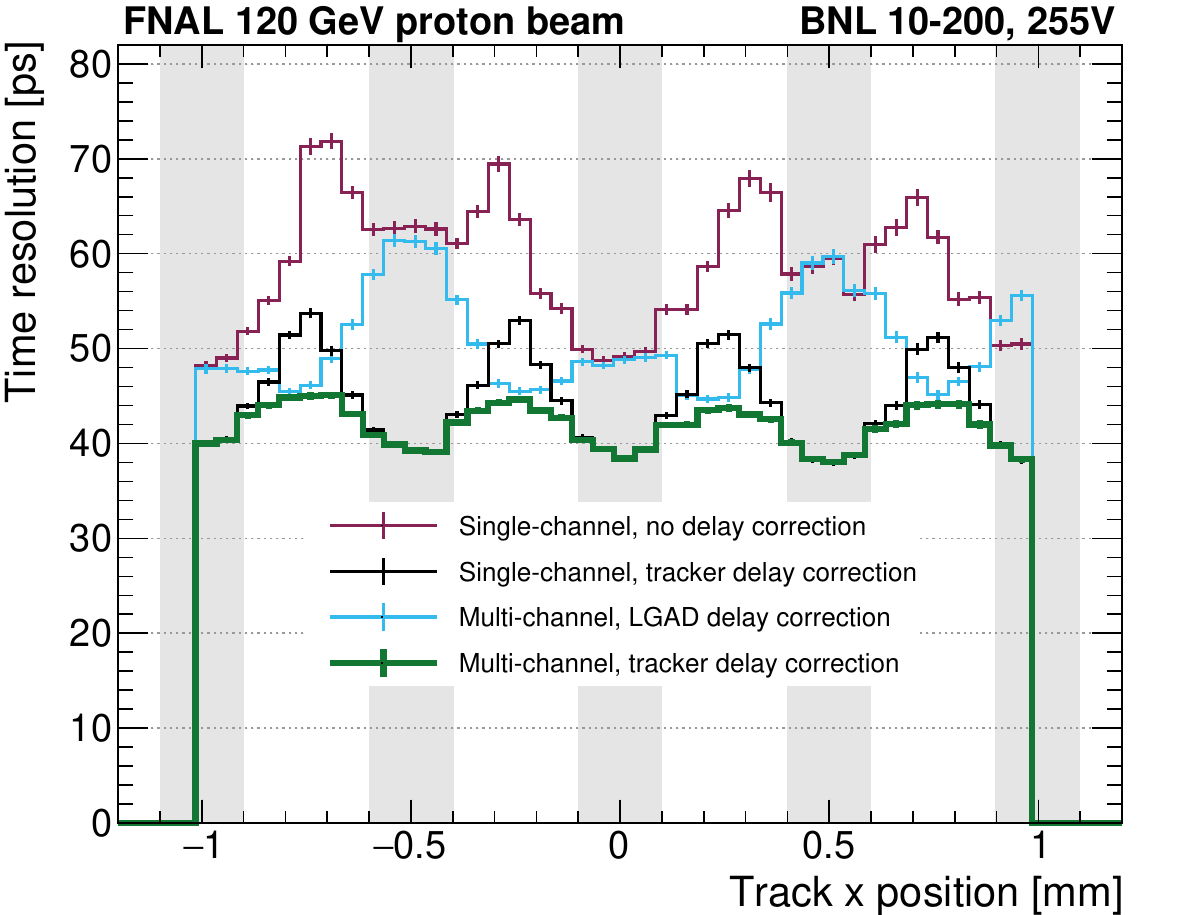}
    \includegraphics[width=0.49\textwidth]{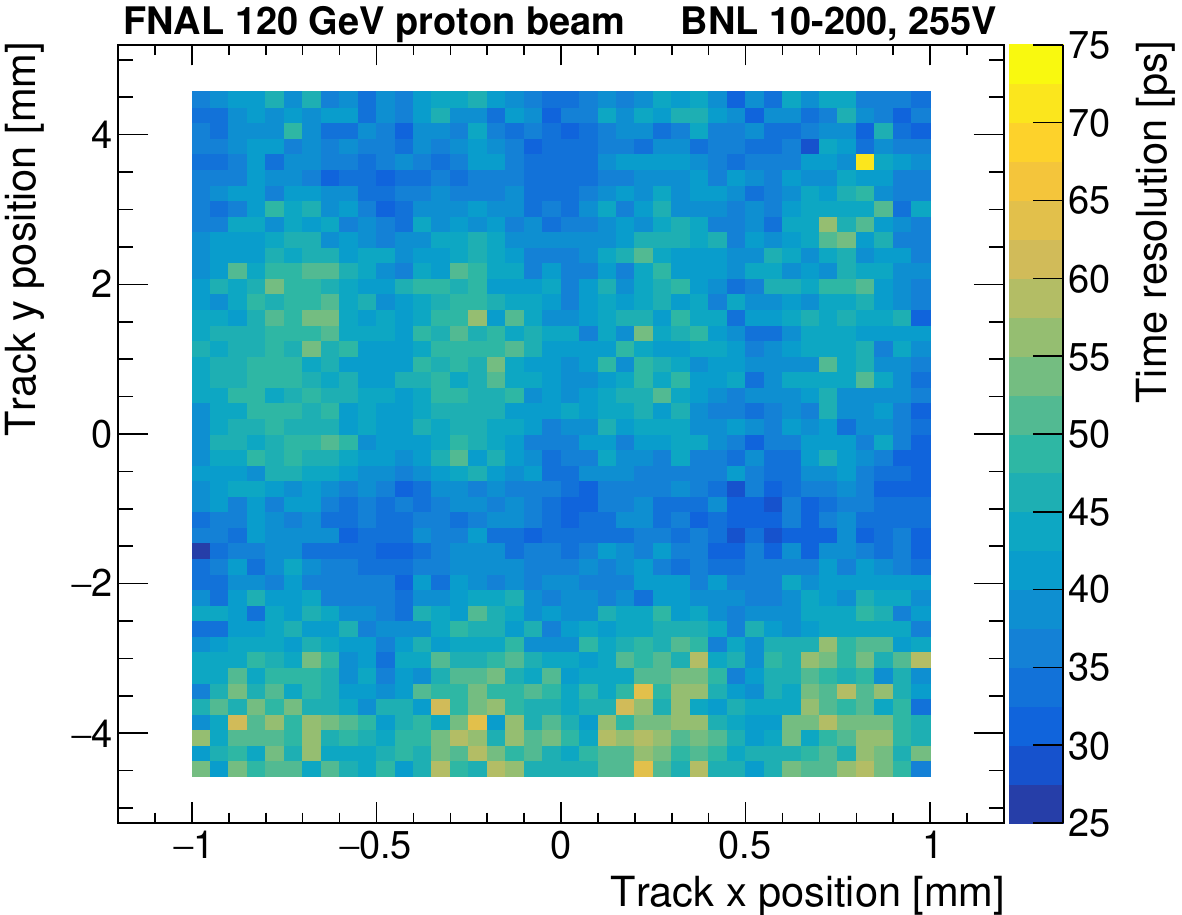}
    \hspace{0.1cm}
    \caption{Time resolution for various time definitions as a function of the telescope track $x$ position for the BNL 10-200 sensor (left).
    The event selection includes hits across the entire sensor surface, including high- and low-gain regions. 
    The four curves show different reconstruction and time delay correction options. The time resolutions shown include single channel reconstruction without a time delay correction, single channel reconstruction with a telescope position dependent time delay correction, multi-channel reconstruction with a telescope position dependent time delay, and multi-channel reconstruction with a position dependence self-corrected using the AC-LGAD $x$ and $y$ position reconstruction. The time resolution as a function of the telescope track $x$ and $y$ position for the multi-channel reconstruction with telescope tracker position time delay correction (right).}
    \label{fig:TimeResolution}
\end{figure}

\begin{figure}[htp]
    \centering
    \includegraphics[width=0.49\textwidth]{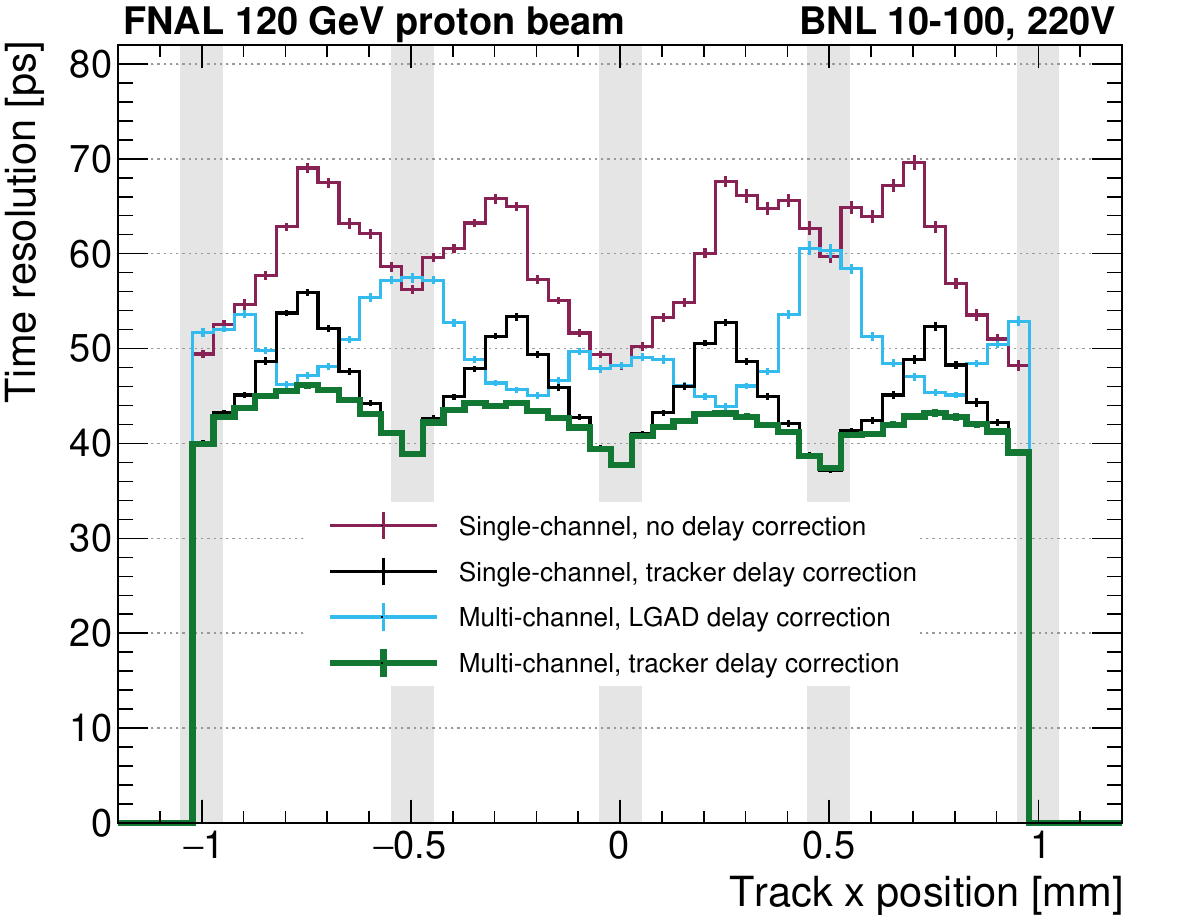}
    \includegraphics[width=0.49\textwidth]{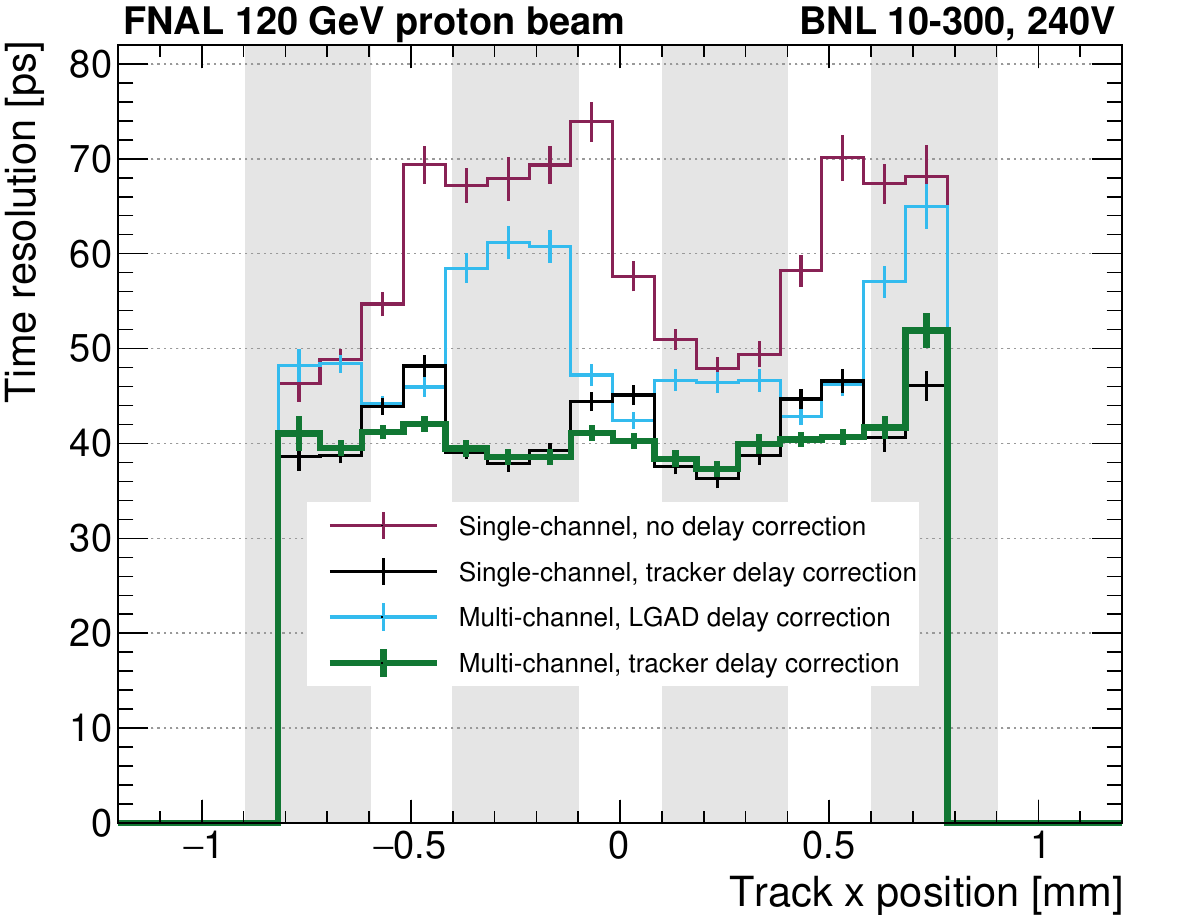}
    \hspace{0.1cm}
    \caption{Time resolution for various time definitions as a function of the telescope track $x$ position for the BNL 10-100 (left) and BNL 10-300 (right) sensors. 
    The event selection includes hits across the entire sensor surface, including high- and low-gain regions. 
    The four curves are based on a variety of time reconstruction and time delay correction definitions, as described in Figure~\ref{fig:TimeResolution}.}
    \label{fig:TimeResolution_Width}
\end{figure}

The time resolution is measured by comparing the AC-LGAD timestamps with those from the Photek MCP-PMT time reference. 
The time resolution of the Photek MCP-PMT has been measured to be 10 ps, and this value is subtracted in quadrature from the time resolution measurements, as discussed in Section~\ref{sec:setup}. 
The time resolutions for the three \SI{10}{\mm} sensors as a function of the proton $x$ impact parameter are shown in Figures~\ref{fig:TimeResolution} and~\ref{fig:TimeResolution_Width} using both the single and multi-channel timestamps and different forms of delay correction, including events along the full length of the strips. 
The purple curves show the single-channel timestamps with no position-dependent delay correction. 
In this case, the uncorrected delays degrade the resolution to \SIrange[range-phrase=--]{50}{70}{\pico\second}. 
Adding the tracked-based delay corrections (black curves) significantly improves the resolution, achieving on average \SI{40}{\pico\second} for direct metal hits, and up to \SI{55}{\pico\second} in the gaps. 
In the gap regions, the signal is shared roughly equally between the two neighbor strips and neither strip has a very large amplitude. 
However, combining both signals into the multi-channel timestamp with delay correction (dark green) recovers most of the performance in the gap regions, yielding average time resolutions of \SIrange[]{40}{45}{\pico\second} for all $x$ positions in all three sensors. 

The absolute time resolution values shown in Figures~\ref{fig:TimeResolution} and~\ref{fig:TimeResolution_Width} are dominated by the sensor regions with low, non-uniform gain, which results in slightly degraded time resolution with respect to the one shown in Figure~\ref{fig:SummaryResolutions200} for the high-gain regions only.
As a result, these resolutions have a non-negligible jitter contribution and do not reach the \SI{30}{\pico\second} Landau limit. 
The impact of the gain non-uniformity can be seen in Figure~\ref{fig:TimeResolution} (right), which shows the multi-channel resolution across the surface of the BNL 10-200 sensor. The spots with high gain (as visible in Figure~\ref{fig:AmplitudeROI_EIC1cm200um}) do achieve \SI{30}{\pico\second} resolution. 
The resolutions considering high-gain regions alone are also shown in Table~\ref{table:Summary}. 
Based on the performance in these regions, we expect similar sensors with uniform, high gain would achieve uniform \SIrange[]{30}{35}{\pico\second} resolution. 

Additionally, we also consider the time resolution using delay corrections based on the self-measured impact parameters in $x$ and $y$ described in Section~\ref{sec:posreco} rather than the external tracker (light blue curves in Figures~\ref{fig:TimeResolution} and~\ref{fig:TimeResolution_Width}). 
In the gap regions, where the position reconstruction performs well, the resolution with AC-LGAD-based delay corrections is very close to what is obtained with the tracker-based correction. 
In the metal strips, however, the majority of events have a signal in only one strip, and the $y$ position reconstruction is not possible. 
In this case, the time resolution degrades to the uncorrected scenario. 

Table~\ref{table:Summary} also summarizes the performance of the shorter and longer sensors. As foreshadowed in Section~\ref{sec:properties}, the slower risetimes in the \SI{25}{\mm} strips lead to significant degradation in the time resolution, even considering only the high gain region. As described in Section~\ref{sec:properties}, we foresee several possible avenues to mitigate this effect and deliver good timing performance even with very long strips.

\section{Conclusions and Outlook}\label{sec:discussion}






In this paper, we presented the results of the first comprehensive beam test campaign studying centimeter-scale AC-LGAD strips. 
The sensors considered were manufactured at the Brookhaven National Laboratory and have a coarse pitch of \SI{500}{\micro\m}. 
They deliver precise 4D measurements over a large area with relatively sparse readout and few channels. 
These traits make these sensors particularly desirable for applications including an Electron Ion Collider 4D tracking layer, or space-based 4D trackers.

\begin{figure}[htp]
    \centering
    \includegraphics[width=0.49\textwidth]{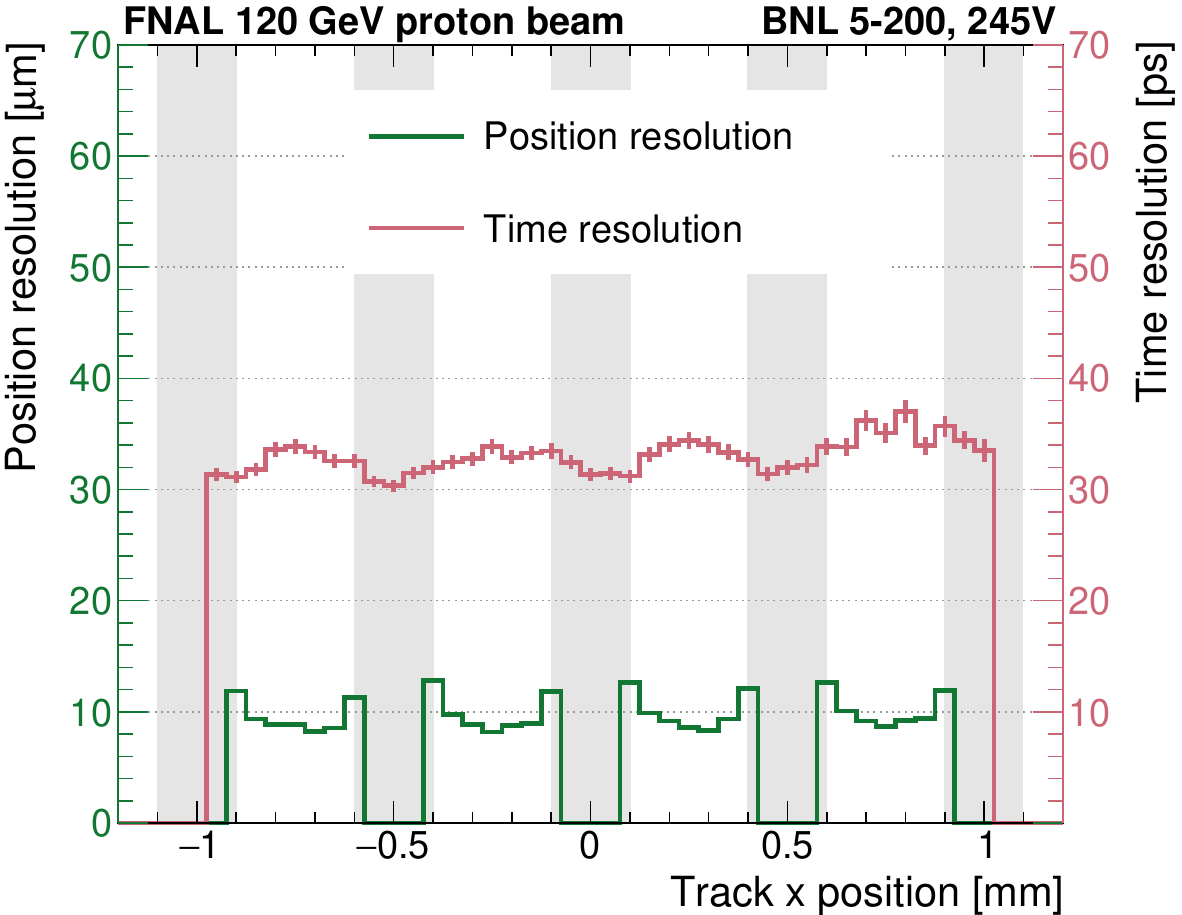}
    \hspace{0.1cm}
    \includegraphics[width=0.49\textwidth]{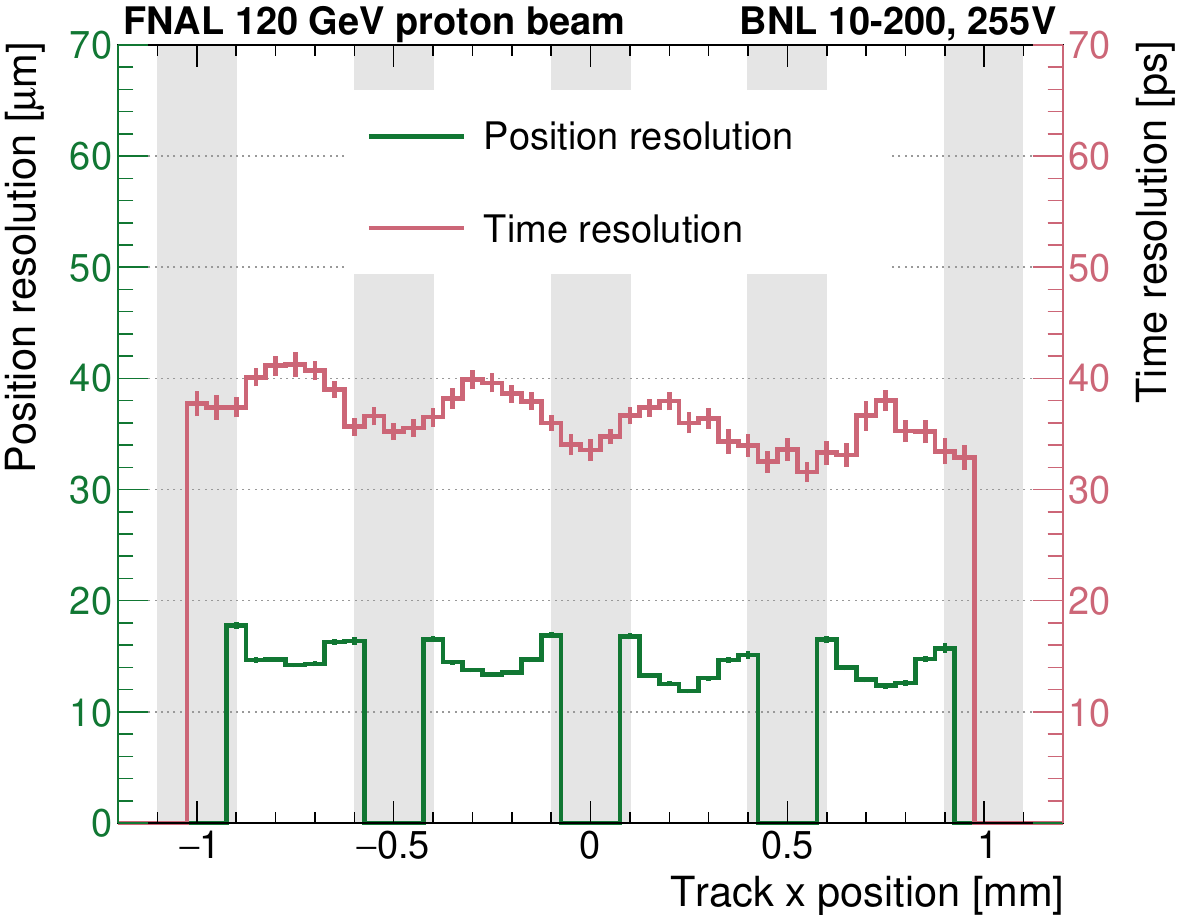}
    \caption{Summary of the position and time resolutions as a function of the track $x$ position for the BNL 5-200 sensor (left) and BNL 10-200 (right). Both the position and time resolutions are measured in the sensor region with high-gain only. The position resolution curve additionally requires at least two strips.}
    \label{fig:SummaryResolutions200}
\end{figure}

We performed a survey of approximately 15 sensors from a single BNL production with varying strip geometries. 
We focused the complete analysis on a set of five sensors spanning the full range of strip lengths and widths in order to understand how to better optimize the electrode design. 
This production suffered from non-uniform gain layer implantation, which degrades the time and spatial resolutions across most of the surface of each sensor. 
However, with careful analysis, we can still extract useful lessons for electrode optimization and estimate the performance that would be obtained in devices with uniform gain. 
The resolutions obtained with the 5 fully analyzed sensors are summarized in Table~\ref{table:Summary}. 
Despite the poor uniformity, most of the sensors still deliver excellent time resolutions in the range of \SIrange[]{30}{50}{\pico\second} and spatial resolutions from \SIrange[]{10}{80}{\micro\m}.
This level of performance as a function track $x$ position can also be seen for the \SI{5}{\mm} (BNL 5--200) and \SI{10}{\mm} (BNL 10-200) long sensors with \SI{200}{\um} metal widths in Figure~\ref{fig:SummaryResolutions200}.

Considering the spatial reconstruction performance, we found in Section~\ref{sec:posreco} that events with clusters of two or more strips convey much more information than single-strip events.
The two-strip clusters allow precise interpolation for the $x$ impact parameter, and to a lesser extent, even the longitudinal impact parameter, $y$, along the strips. 
As particles that pass directly through a metal strip tend to yield signals completely absorbed in that strip, this presents a strong motivation to minimize the electrode width and maximize the acceptance for two-strip clusters. 
In turn, reducing the electrode width confines the remaining single-strip events to the small region within the strips, reducing the spatial resolution of this category to approach that of the two-strip category. 
Based on the performance observed, we believe strips of \SI{500}{\micro\m} pitch and \SI{100}{\micro\m} electrode width and uniform gain could attain uniform spatial resolution of \SIrange[]{15}{20}{\micro\m}. We extrapolate that strips with even narrower \SI{50}{\micro\m} width may yield better performance, constraining the single-strip population to a smaller region.
Furthermore, the resistivity of the n+ layer could also be tuned to ensure more uniform two-strip acceptance.

On the other hand, precision timing analysis with such large devices presents interesting features not observed in shorter AC- or DC-LGADs considered in previous studies. 
Signals propagating long distances across the sensor surface are collected with varying delays at a scale at least as large as the desired time resolution. 
Since these delays are determined by the impact position, they can be easily corrected with a measurement of hit location with modest resolution. 
Within a collider application, the tracking detector performance is typically far in excess of what is needed for such a correction to make the impact of the position-dependent delays negligible. 
Even for operation as a standalone timing plane or in real-time operation at a collider before tracking is performed, the 2D spatial reconstruction is sufficient to remove the effect of the delays. 
This consideration, too, strongly favors thinner electrodes and expanded two-strip cluster acceptance, which makes the longitudinal reconstruction and standalone position-dependent delay correction possible.

The other interesting challenge for the time resolution is related to the impact of the strip length, as described in Section~\ref{sec:properties}. 
The longest sensors with strip lengths beyond \SI{2}{\centi\m} exhibit slower rising edges that increase the susceptibility to jitter in the timing measurement. 
We outlined several potential mitigation techniques, and it is likely possible to design \SI{2.5}{\centi\m} strips with good timing properties. 
In particular, reducing the electrode width, as already motivated by the spatial reconstruction, would reduce the electrode capacitance and may be enough to restore the fast rising edge even in the longest strips. 
These options will be explored in the next beam campaign. 
The strip length is less relevant for the spatial reconstruction. Although the longest prototype analyzed in this campaign suffered from poor gain uniformity and yielded poor performance, we do not find any indication that the strip length itself should necessarily have a large impact on the spatial resolution. Any modification of the pulse shape due to the length could be mitigated using a measure of the pulse integral, rather than amplitude, for the position reconstruction.

Taking all of the results together, we believe that large area, coarse pitch sensors with uniform \SI{30}{\pico\second} and \SIrange[]{15}{20}{\micro\m} resolutions are well within reach. 
This performance can be achieved with a moderate optimization of the layout and processing compared to the analyzed sensors, including reducing the strip width and the surface resistivity. 
Improving the gain uniformity is an important consideration as well, but uniformity at this scale is routinely delivered by HPK and FBK for the CMS and ATLAS timing detectors, and is expected to be achieved in future BNL productions, too. 
Finally, although strips up to \SI{1}{\centi\m} have already demonstrated signal shapes adequate for \SI{30}{\pico\second} resolution, maintaining this performance beyond \SI{2}{\centi\m} require further study and confirmation in the next test beam campaign.

As shown in Table~\ref{table:Summary} and Figure~\ref{fig:SummaryResolutions200}, the BNL 10--200 sensor can achieve simultaneous \SI{32}{\pico\second} and \SI{18}{\micro\m} resolutions considering events from the high gain regions with at least two strips. 
As result, this class of sensors represents a strong candidate for future 4D tracking detectors, providing excellent position and time resolutions with small channel count compared to standard tracking sensors.

\begin{table}[htp]
  \centering
  \caption{Summary of spatial and timing performance for the five fully analyzed sensors. 
  The time resolution for the high gain regions is shown. 
  The position resolutions and efficiencies are shown for both the one and two strip categories. 
  }
  \begin{tabular}{ l | c || c | c | c | c }
  \multirow{5}{*}{Name} & \multicolumn{1}{c||}{Time resolution} & \multicolumn{4}{c}{Spatial resolution} \\
  \cline{2-6}
  & & \multicolumn{2}{c|}{Exactly one strip} & \multicolumn{2}{c}{Two strip} \\
  \cline{3-6}
  & High gain & Resolution & Eff. & Resolution & Eff. \\
             
  Unit            & \si{ps} & \si{\um} & - & \si{\um} & - \\
  \hline\hline
  BNL  5--200     & 30 $\pm$ 1 & 61 $\pm$ 1 & 35\% & 12 $\pm$ 1 & 65\% \\ \hline
  BNL 10--100     & 35 $\pm$ 1 & 69 $\pm$ 1 & 23\% & 19 $\pm$ 1 & 77\% \\
  BNL 10--200     & 32 $\pm$ 1 & 82 $\pm$ 1 & 43\% & 18 $\pm$ 1 & 57\% \\
  BNL 10--300     & 36 $\pm$ 1 & 83 $\pm$ 1 & 51\% & 16 $\pm$ 1 & 49\% \\ \hline
  BNL 25--200     & 51 $\pm$ 1 &128 $\pm$ 1 & 82\% & 31 $\pm$ 1 & 18\% \\
  \end{tabular}
  \label{table:Summary}
\end{table} 

\acknowledgments
We thank the Fermilab accelerator and FTBF personnel for the excellent performance of the accelerator and support of the test beam facility, in particular M.~Kiburg, E.~Niner, N.~Pastika, E.~Schmidt and T.~Nebel. 
We also thank the SiDet department, in particular M.~Jonas and H.~Gonzalez, for preparing the readout boards by mounting and wirebonding the AC-LGAD sensors. 
Finally, we thank L.~Uplegger for developing and maintaining the telescope tracking system, and R.~Lipton for useful discussions.

This document was prepared using the resources of the Fermi National Accelerator Laboratory (Fermilab), a U.S. Department of Energy, Office of Science, HEP User Facility. 
Fermilab is managed by Fermi Research Alliance, LLC (FRA), acting under Contract No. DE-AC02-07CH11359.
This work was also supported by the U.S. Department of Energy under grant DE-SC0012704; 
used resources of the Center for Functional Nanomaterials, which is a U.S. DOE Office of Science Facility, at Brookhaven National Laboratory under Contract No. DE-SC0012704;
supported by the Chilean ANID PIA/APOYO AFB180002 and ANID - Millennium Science Foundation - ICN2019\_044.
This research was also performed using the resources of the University of Illinois at Chicago and supported by the U.S. Department of Energy under grant DE-FG02-94ER40865.

\bibliographystyle{report}
\bibliography{biblio}{}
\end{document}